\documentclass[apj]{emulateapj}

\usepackage{apjfonts}
\usepackage{epsfig}
\usepackage{graphicx}
\usepackage{url}
\usepackage{natbib}
\usepackage{float}
\usepackage{amsmath}
\usepackage{color}
\usepackage[backref,breaklinks,colorlinks,citecolor=blue]{hyperref}
\usepackage[all]{hypcap}

\begin{document}

\shorttitle{SHARDS of Disrupted Milky Way Satellites with LAMOST}
\shortauthors{Carlin et al.}

\title{Characterizing the SHARDS of Disrupted Milky Way Satellites with LAMOST}

\author{
Jeffrey L. Carlin\altaffilmark{1,2}, 
Chao Liu\altaffilmark{3},
Heidi Jo Newberg\altaffilmark{1},
Timothy C. Beers\altaffilmark{4},
Licai Deng\altaffilmark{3},  
Puragra Guhathakurta\altaffilmark{5},
Zihuang Cao\altaffilmark{3},
Yonghui Hou\altaffilmark{6},
Yuefei Wang\altaffilmark{6},
Yong Zhang\altaffilmark{6}
}

\altaffiltext{1}{Department of Physics, Applied Physics and Astronomy, Rensselaer Polytechnic Institute, Troy, NY 12180, USA, jeffreylcarlin@gmail.com}
\altaffiltext{2}{Department of Physics and Astronomy, Earlham College, Richmond, IN 47374, USA}
\altaffiltext{3}{Key Lab of Optical Astronomy, National Astronomical Observatories, Chinese Academy of Sciences, Beijing 100012, China}
\altaffiltext{4}{Department of Physics and JINA Center for the Evolution of the Elements, University of Notre Dame, Notre Dame, IN 46556, USA}%
\altaffiltext{5}{UCO/Lick Observatory, Department of Astronomy and Astrophysics, University of California, Santa Cruz, CA 95064, USA}
\altaffiltext{6}{Nanjing Institute of Astronomical Optics \& Technology, National Astronomical Observatories, Chinese Academy of Sciences, Nanjing 210042, China}

\begin{abstract}

We derive the fraction of substructure in the Galactic halo using a
sample of over 10,000 spectroscopically-confirmed halo giant stars
from the LAMOST spectroscopic survey. By observing 100 synthetic
models along each line of sight with the LAMOST selection function in
that sky area, we statistically characterize the expected halo
populations. We define as SHARDS (Stellar Halo Accretion Related
Debris Structures) any stars in $>3\sigma$ excesses above the model
predictions. We find that at least $10\%$ of the Milky Way halo stars
from LAMOST are part of SHARDS.  By running our algorithm on smooth
halos observed with the LAMOST selection function, we show that the
LAMOST data contain excess substructure over all Galactocentric radii
$R_{\rm GC} < 40$~kpc, beyond what is expected due to statistical
fluctuations and incomplete sampling of a smooth halo. The level of
substructure is consistent with the fraction of stars in SHARDS in
model halos created entirely from accreted satellites.  This work
illustrates the potential of vast spectroscopic surveys with high
filling factors over large sky areas to recreate the merging history
of the Milky Way.

\end{abstract}


\section{Introduction}

Though the stellar halo of the Milky Way (MW) contains only $\sim1\%$
of the Galaxy's total stellar mass, long dynamical times in the halo
mean that much of the fossil record of the formation and evolution of
our Galaxy is preserved in dynamical and chemical signatures of halo
stellar populations. In the prevailing $\Lambda$-Cold Dark Matter
($\Lambda$CDM) cosmological paradigm, massive galaxies grow by the
agglomeration of many smaller sub-galactic fragments, or subhalos,
that contribute their constituent dark matter, stars, and gas to the
larger host galaxy (e.g., \citealt{wr78,ws00}). Through detailed
analysis of the kinematics and chemistry of stellar populations in the
MW halo, we can thus assess the relative importance of the monolithic
collapse suggested by \citet{els62} as a possible origin for the
stellar halo, and the stochastic process of accretion as discussed by,
e.g., \citet{sz78}.

Deep, large-area photometric surveys such as the Two Micron All-Sky
Survey (2MASS; \citealt{scs+06}) and the Sloan Digital Sky Survey
(SDSS; \citealt{yaa+00}) have laid bare the ubiquitous substructure
lurking in the low surface brightness stellar halo. The numerous
substructures that have been discovered as stellar overdensities
include the Sagittarius (Sgr) tidal stream (e.g., \citealt{igi94,
msw+03, bze+06, kbe+12}), the large, complex overdensity in Virgo
(e.g., \citealt{vza+01, nyc+07, cyc+12, dvz+14}), the
Triangulum-Andromeda (TriAnd) cloud (e.g., \citealt{mor+04, rms+04,
mii07, sjm+14, pjs+15}), and the Orphan Stream (e.g., \citealt{g06,
bei+07, nwy+10}), among others (for a review of currently known
structures, see \citealt{gc16}).

It is thus becoming clear that not only is much of the outer Galactic
halo made up of accreted debris, but that this is an ongoing
process. Tidal remnants that are visible as number-density
enhancements are only the most recent infall events, and do not
constitute a representative sample of satellite accretion at all
epochs of MW evolution. Fortunately, accretion remnants remain
relatively coherent in phase space over much longer time scales (e.g.,
\citealt{hw99}), making it possible to identify relics of infall
events dating back many Gyr ago (note, however, that debris from the
most ancient accretion events will have phase-mixed, and become
indistinguishable from a smooth halo at present day; see, e.g.,
\citealt{jbs+08}). In addition, the long-lived low-mass stars from
accreted dwarf galaxies and star clusters retain the chemical
signatures of their parent satellite at the time of their
formation. It is thus possible to recreate a portion of the
hierarchical formation history of the Galaxy by detailed analysis of
the stellar halo, by which one can determine the fraction of the halo
that resides in substructure at present, and use this to assess the
relative contribution of in situ star formation (i.e., stars formed
from gas residing in the deep potential well of the MW) and accretion
(stars formed in dwarf galaxy or stellar cluster potentials).

Full cosmological models of Milky Way-mass galaxies with sufficient
resolution to include all the relevant physics at all physical scales
are challenging.\footnote{Progress is being made, however; see, e.g.,
\citealt{tbc+14} and references therein.} However, with the
combination of cosmologically-motivated models and semi-analytic
techniques, robust predictions have been made to guide our intuition
about the formation of the Galactic halo. \citet[hereafter
``BJ05'']{bj05} created model stellar halos from a suite of 11
$N$-body models of satellites from cosmologically-motivated merger
trees, to which stellar populations were assigned in a manner that
reasonably matches the properties of MW satellites. Because these
halos are generated purely from accreted satellites, they offer an
extreme case to compare with our MW halo data. Though they are created
entirely from satellites, these halos contain a smooth population of
older debris in the inner halo (inside $R_{\rm GC}\sim20$~kpc), with
more recent, unmixed accretion events predominating in the outer
halo. Analysis of these halos by \citet{jbs+08} showed that different
types of structures (great circles, clouds, and mixed remnants) result
from different families of satellite orbits. Substructure dominates in
the mock halos at large radii, but is also biased toward cloud-like
remnants, which result from disruptions on radial
orbits. \citet{jbs+08} found that the most recently accreted
substructures should be more metal-rich than the smooth halo (which
consists predominantly of metal-poor, phase-mixed early
debris). Finally, these authors concluded that on average $\sim10\%$
(ranging from $1-50\%$) of stars should be currently visible in
substructure, with the fraction depending on the epoch of accretion.

\citet{ans06} used a suite of eight high-resolution cosmological ($N$-body+hydrodynamic) simulations of Milky Way-like galaxies to explore the outer stellar halo. They found that outside $\sim20$~kpc, $\sim95\%$ of the stars are accreted; the in situ stars found outside this radius were kicked up in merger events. This leads to a ``break'' in the radial surface brightness profile (SBP), where there is excess luminosity beyond the break radius compared to an
extrapolation of the inner SBP. This steepening of the mass profile
with radius was shown to be consistent with the spatial distribution
of MW and M31 outer-halo globular clusters, suggesting that they also
originated in accretion events. \citet{zwb+09} used high-resolution
cosmological $N$-body+SPH simulations of 4 MW-like galaxies to assess
the fraction of in situ halo stars. In the highest-resolution of their
simulations, $\sim20\%$ of the halo stars were formed in situ, and are
currently located within $\sim20-30$~kpc of the galaxy center. These
in situ stars form deep in the potential well of the host, and are
kicked out via merger events. \citet{zwb+09} found that the fraction
of in situ halo stars is higher ($20-50\%$) in the inner halos of
hosts with relatively quiescent recent merger histories compared to
those with more recent merger activity.

In addition to the detection and characterization of high number
density, coherent halo substructures resulting from recent tidal
disruption, recreating the merger history of the Milky Way requires
systematic searches for kinematically cold structures from older
accretion events. \citet{sra+09} searched for these ``Elements of Cold
HalO Substructure (ECHOS)'' in 137 individual SDSS lines of sight,
each covering $\sim7$ square degrees on the sky. Based on identified
statistically significant radial velocity peaks among inner-halo
main-sequence turnoff (MSTO) stars (at distances of 10-17.5~kpc from
the Sun), these authors estimated that an upper limit of $\sim1/3$ of
the metal-poor MSTO stars in the inner halo reside in
ECHOS. \citet{shm+09} developed a metric (the {\it 4distance}) for
quantifying the separation of stars in 4-dimensional position-velocity
space. Their application of the {\it 4distance} to the 101 halo giants
from the Spaghetti survey \citep{mmo+00} found that 20 stars
($\sim20\%$ of the sample) are in groups, compared to an expectation
of 9 from a randomized sample. Thus the authors place a lower limit
that more than 10\% of halo stars must be in substructure. The {\it
4distance} metric was also applied to a sample of $\sim4000$ BHB stars
from SDSS by \citet{xry+11}, who showed that the observations have
significantly more {\it 4distance} ``pairs'' than a smooth halo, but
are deficient in substructure when compared to the pure-accretion
halos of \citet{bj05}. This is, however, reconciled when comparing
only the old stars from the models; thus, the (old, metal-poor) BHB
stars must only be tracing a small fraction of the predominantly
metal-rich, recent-infall debris. Most recently, the {\it 4distance}
metric was also applied to a sample of 4568 SEGUE K~giants by
\citet{jmm+16}. By comparing the number of groups (created via
friends-of-friends aggregation of pairs identified by the {\it
4distance}) to a smooth model halo, this work showed that $\sim50\%$
of the halo stars in their sample are identified with groups (though
with significant contamination from false positives). The fraction of
stars in substructure increases with Galactocentric radius, and is
also higher in more metal-rich populations than in the most metal-poor
halo stars. A large fraction ($>50\%$ beyond Galactocentric radius of
30 kpc) of the groups are noted by \citet{jmm+16} to likely be
associated with the Sgr stream. A correlation function statistic was
applied by \citet{ccf+11} to the SDSS BHB star sample from
\citet{xrz+08} to search for spatial and kinematic correlations in the
Galactic halo. The number of BHB stars was found to be deficient at
$R_{\rm GC}>30$~kpc compared to the predictions of the Aquarius
simulations of MW-like halos, though significant clustering is present
in the BHB sample at these large radii that is consistent with model
halo predictions.

Statistical characterization of the fraction of substructure in the
halo need not be limited to spectroscopically-identified
samples. \citet{bzb+08} used SDSS DR5 photometry of MSTO stars to fit
the density profile of the Galactic halo. Based on the rms residuals
about this fit, this work concluded that $> 40\%$ of the halo is in
substructure. This fraction increases with distance, nearly doubling
from 10-35 kpc, though much of this rise can be attributed to Sgr and
other known substructures. Bell et al. showed that the measured rms is
consistent with some of the \citet{bj05} models, which lends support
to the notion that the MW halo is largely accreted debris.
\citet{hcw+11} analyzed Aquarius models (with stellar populations from the semi-analytic method of \citealt{ccf+10}), and found that the rms stellar density in halos with no smooth component is much larger than that observed in SDSS by \citet{bzb+08}. Addition of a 10\% smooth
halo component brings them roughly into agreement, though because the model halos are highly anisotropic, the vantage point of the observer affects this significantly. 

In this work, we measure the fraction of substructure among Galactic
halo stars observed by the LAMOST spectroscopic survey. Our method is
similar to the {\it 4distance} technique developed by \citet{shm+09},
but because the LAMOST depth is not uniform along different lines of
sight, we choose not to use a global correlation statistic, but
instead identify statistical excesses in separate regions of
sky. LAMOST will eventually obtain a nearly complete magnitude-limited
sample (at high Galactic latitudes) over a huge contiguous region of
the northern sky. Targets for the LAMOST survey are selected with a
smoothly-varying selection function, making the sample a relatively
unbiased data set. The combination of a vast contiguous sky area and
the simple selection function make this a valuable data set for
identifying substructures on a variety of spatial scales. Though the
selection function for a given direction is simple, we nevertheless
must model its effects over the entire survey; to do so, we compare
all of our results in this work to mock observations in which we apply
the LAMOST selection function to the Galaxia model \citep{sbj+11} in
each region of sky.  Finally, we compare our results to
accretion-derived halos from \citet{bj05}, again observed with the
LAMOST selection function.

This paper is outlined as follows. In Section 2, we discuss the LAMOST
survey and the selection of Galactic halo stars for our substructure
search. Section 3 outlines our technique for identifying
substructures, which we dub SHARDS (Stellar Halo Accretion Related
Debris Structures). In Section 4 we compare our results to the Galaxia
model of the smooth halo, and to halos built entirely from disrupted
satellites. We obtain estimates for the fraction of halo stars in
substructure in Section 5, and characterize the overall properties of
the substructures we have identified. We conclude in Section 6 with
some context for our results and a discussion of upcoming work that
builds upon these results.

\section{Available LAMOST data set and its properties}\label{sec:data}

\begin{figure}[!t]
\includegraphics[width=0.5\textwidth, trim=2.5cm 0cm 0cm 0cm, clip]{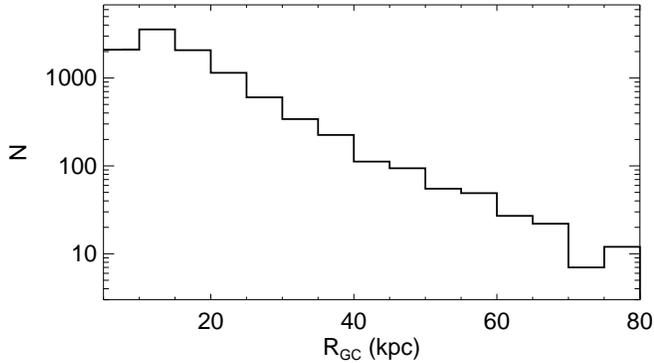}
\caption{Distribution of Galactocentric distances for halo giants in the LAMOST DR1-3 sample. There are 10,481 stars with spectra having derived parameters $S/N>5$, $0 < \log{g} < 3.5$, $3800 < T_{\rm eff} < 6500$~K, and $|Z| > 5$~kpc.}
\label{fig:rgc}
\end{figure}

The Large Sky Area Multi-Object Fiber Spectroscopic Telescope (LAMOST)
survey is an ongoing effort being carried out with the $\sim4$-meter
effective aperture Guoshoujing Telescope in northern China
\citep{czc+12, zzc+12}. The telescope has 4000 robotically-positioned
optical fibers arrayed over a $5^\circ$ diameter focal plane, feeding
16 optical spectrographs that produce spectra with resolution
$R\sim2000$ covering wavelengths $\sim3800 < \lambda <
9000$~\AA. LAMOST has completed 3 years of survey operations plus a
Pilot Survey, and has internally released a total of $\sim5.7$~million
spectra to the collaboration, spanning observation times from
Oct. 2011 -- May 2015.\footnote{The first public data release (LAMOST
DR1; \citealt{lzz+15}) is available at \url{http://dr1.lamost.org/}.}
Of these, $\sim3.1$~million are AFGK-type stars (mostly part of the
LEGUE survey of Galactic structure; \citealt{dnl+12}) with estimated
stellar parameters ($T_{\rm eff}$, $\log{g}$, [Fe/H];
\citealt{wdl+14}), and all objects have available radial velocities or
redshifts accurate to 5-10~km~s$^{-1}$ (e.g., \citealt{lzz+15,
xly+15}). The survey reaches a limiting magnitude of $r = 17.8$ (where
$r$ denotes magnitudes in the SDSS $r$-band), with most targets
brighter than $r\sim17$; LAMOST will achieve a nearly
magnitude-limited survey to $r\sim17$ over much of the high Galactic
latitude northern sky.

\begin{figure}[!t]
\includegraphics[width=0.5\textwidth]{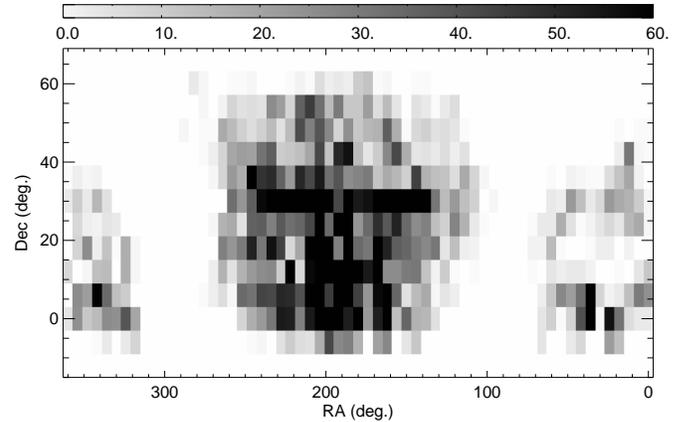}
\caption{Spatial density (in $6^\circ \times 6^\circ$ regions) of the 10,481 halo giants in the LAMOST DR1-3 sample. Note that the colorbar saturates at 60; some bins exceed this total.}
\label{fig:radec_giants}
\end{figure}

To select a sample of Galactic halo giants, we use stellar distance
estimates derived from the LAMOST stellar parameters \citep{cln+15}. This work showed that, given the typical uncertainties in stellar parameters given by the LAMOST pipeline, distances to most stars can be derived to $\lesssim30\%$ accuracy.
Our sample of halo giants is selected from the LAMOST database with
the constraints: $0.0 < (J-K)_0 < 2.0$, $K_0 < 15.5$, spectra having
$S/N>5$ (in either the $g$ or $r$ band), derived parameters $0 <
\log{g} < 3.5$, $3800 < T_{\rm eff} < 6500$~K, and $|Z| >
5$~kpc.\footnote{Throughout this work, we assume the Sun is at ${X, Y,
Z}=(-8, 0, 0)$~kpc in a right-handed Galactic coordinate
system. Line-of-sight velocities are converted to a Galactocentric
frame (i.e., $V_{\rm GSR}$) using a circular velocity of
220~km~s$^{-1}$ and Solar peculiar velocity of $(U, V, W) = (9, 12,
7)$~km~s$^{-1}$.} The color and magnitude cuts exclude stars with
spurious measurements or extreme stellar types (whose distances are
not well-determined). The $|Z|$ criterion ensures that the stars in
our sample are many scale heights beyond the disk plane (the disk
scale height is $\sim 1$~kpc; e.g., \citealt{rhc+96, css+01, smr+02,
jib+08}), minimizing the contribution of disk stars to our sample. We
are interested in assembly of a sample of well-understood,
first-ascent red giant branch (RGB) stars, so the $\log{g}$ and
$T_{\rm eff}$ cuts are intended to remove objects that could be
evolved stars (e.g., RR Lyrae, horizontal-branch, and asymptotic giant
branch stars), while including sufficiently warm stars to retain the
most metal-poor ``normal'' giants. Finally, the $S/N$ criterion, while
generously allowing relatively low $S/N$ spectra into the sample, is
still likely removing many legitimate K-giant stars from our sample;
\citet{ldc+14} showed that K-giants can be reliably identified from
LAMOST spectra even at low $S/N$. For the current study, we choose to
avoid including dwarf contamination that is inevitable in the low
$S/N$ K-giant sample, but will include these in future expansions of
this work. The distribution of Galactocentric distances ($R_{\rm GC}$)
for this sample of 10,481 halo RGB stars is shown in
Figure~\ref{fig:rgc}, and their density on the sky is shown in
Figure~\ref{fig:radec_giants}. The sample contains stars extending
beyond $R_{\rm GC} > 60$~kpc in the halo, with a substantial number of
stars between $40<R_{\rm GC}<60$~kpc. In
Figure~\ref{fig:radec_giants}, most of the stars are at high Galactic
latitudes well away from the plane (as enforced by our $|Z| > 5$~kpc
cut), with nearly the entire north Galactic cap region sampled
substantially by LAMOST.


\section{Stellar excess-finding technique}\label{sec:technique}

We wish to identify substructures that we will refer to as SHARDS
(Stellar Halo Accretion Related Debris Structures). By this we mean
structures that constitute statistical excesses in velocity-distance
space (if we consider localized sky areas, this is essentially a
4-dimensional phase space consisting of 3-D position and line-of-sight
velocity).

Our goal is to assess the fraction of Galactic halo stars observed by
LAMOST that are part of substructures. To do so, we wish to consider
not only structures that are clustered spatially on the sky, but also
those that clump in velocity and distance. Accretion relics remain
coherent in phase space for a longer time than they are visible in
configuration space, so using velocity-distance metrics to identify
structures should sample more of the accretion history of the Milky
Way than simply using spatial clustering and/or distance. Furthermore,
the large areal coverage of the LAMOST survey allows us to probe much
larger spatial scales than surveys that are limited to localized
regions of sky.

\begin{figure*}[!t]
\includegraphics[width=0.33\textwidth, trim=0.3in 0.3in 0.25in 0.3in, clip]{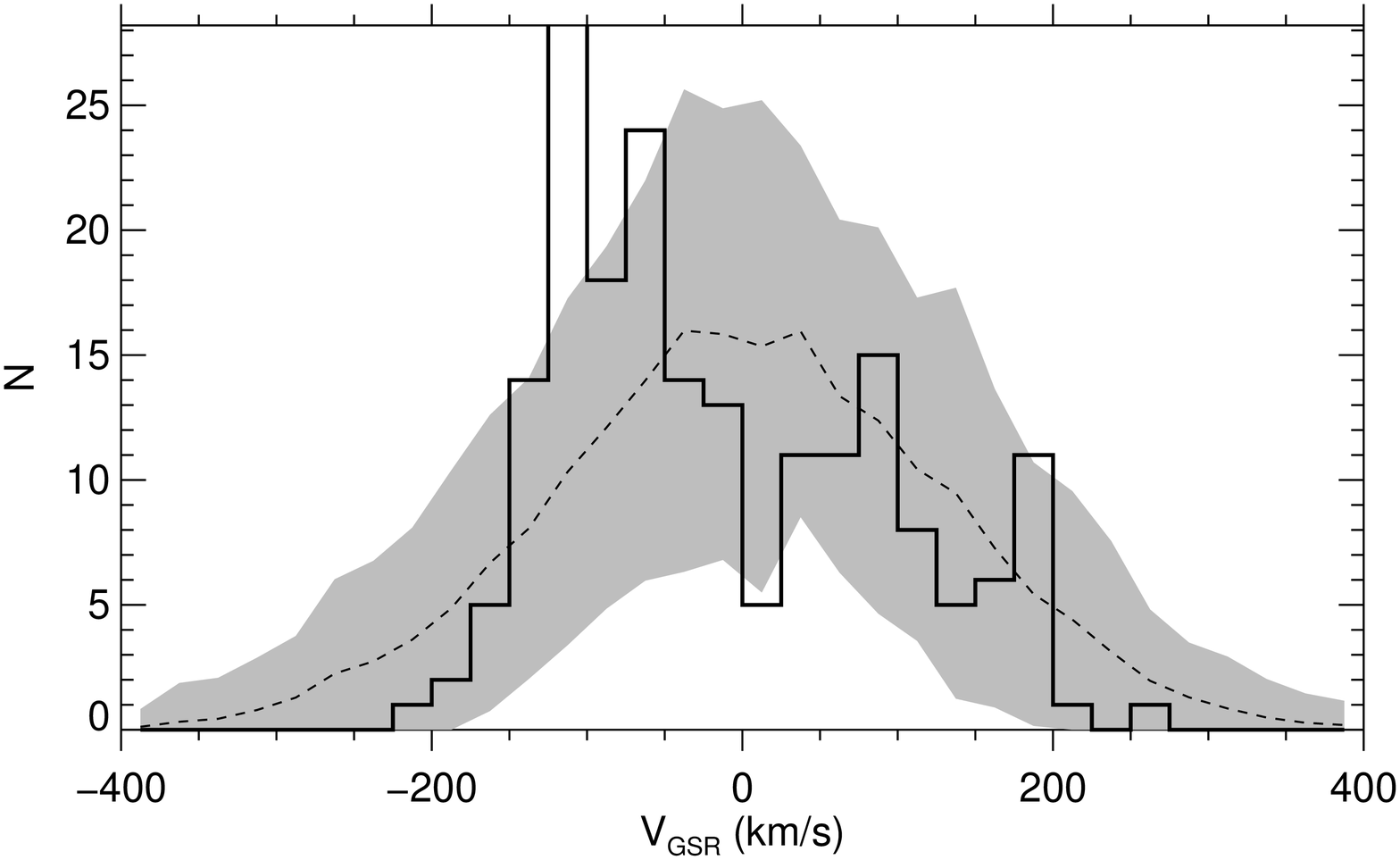}
\includegraphics[width=0.33\textwidth, trim=0.3in 0.3in 0.5in 0.3in, clip]{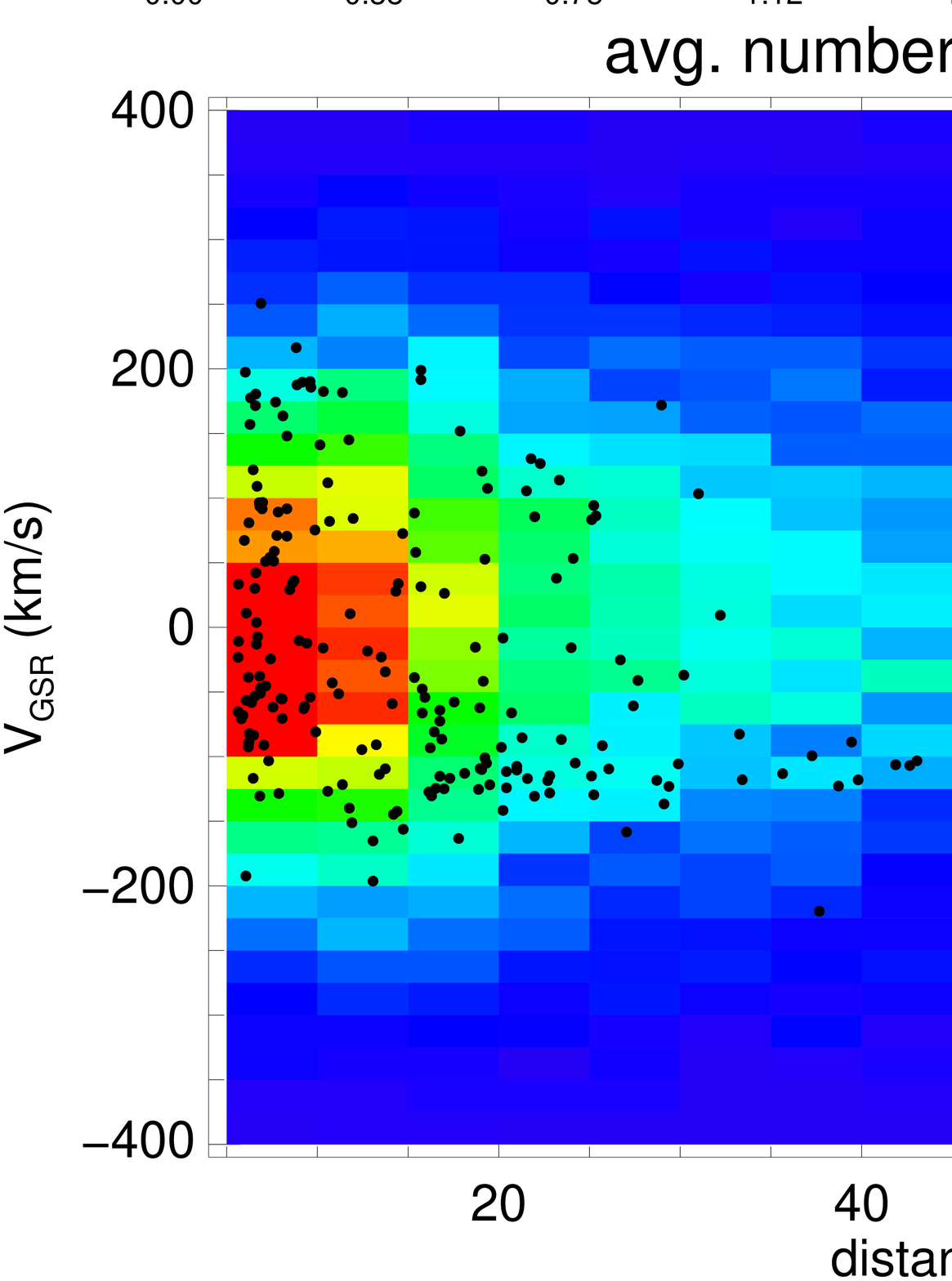}
\includegraphics[width=0.33\textwidth, trim=0.3in 0.3in 0.5in 0.3in, clip]{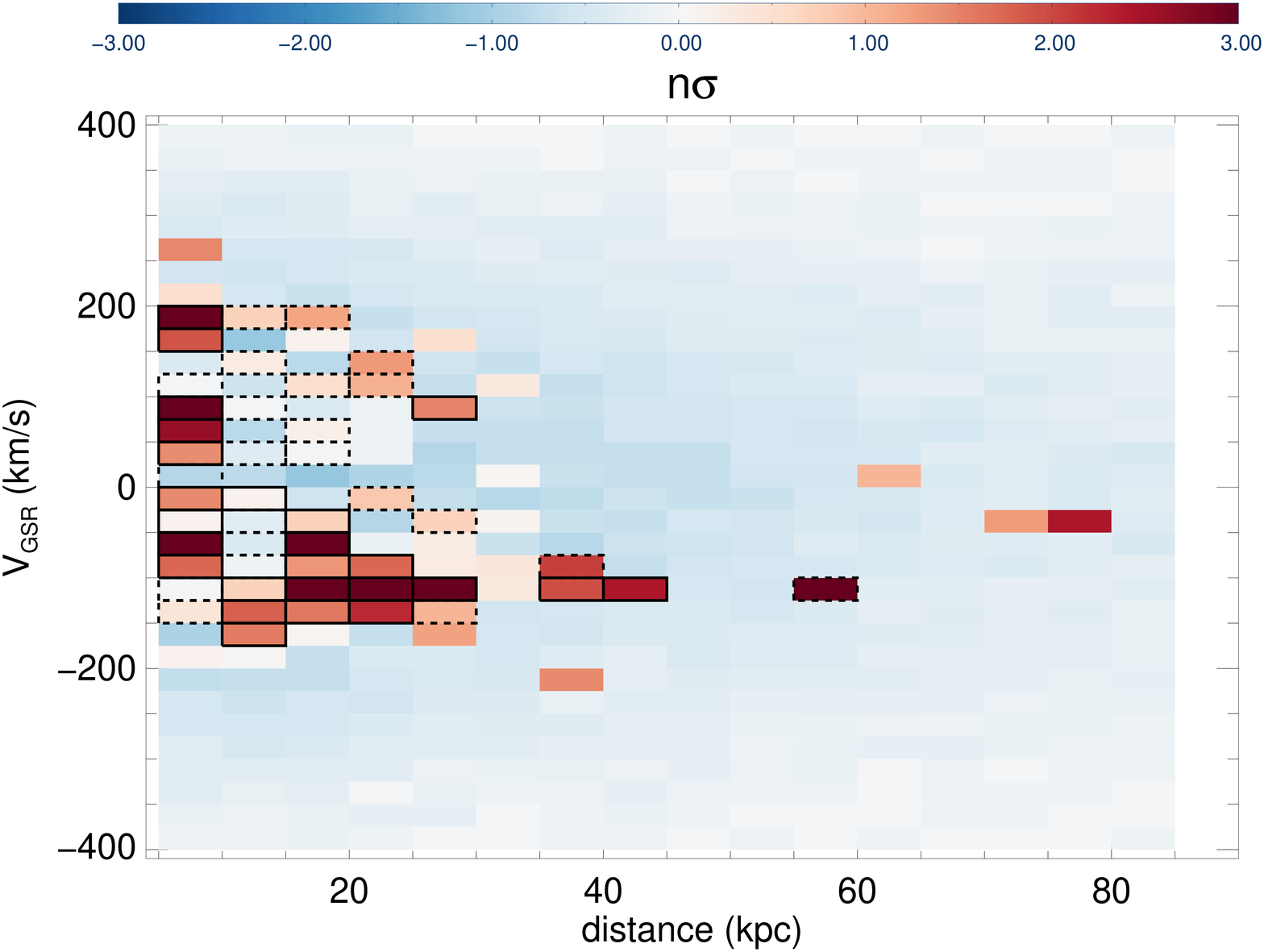}\\
\includegraphics[width=0.33\textwidth, trim=0.3in 0.3in 0.25in 0.3in, clip]{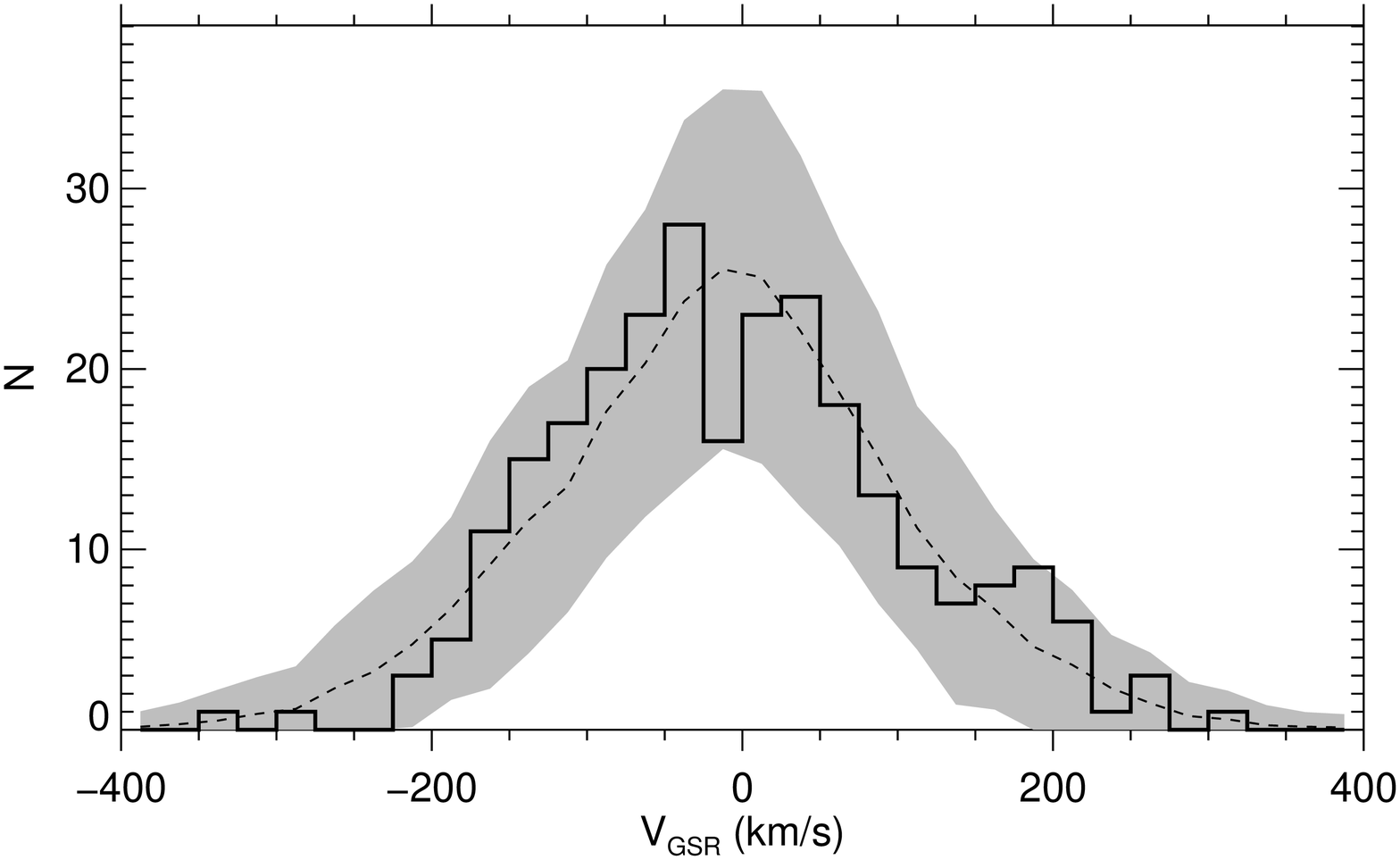}
\includegraphics[width=0.33\textwidth, trim=0.3in 0.3in 0.5in 0.3in, clip]{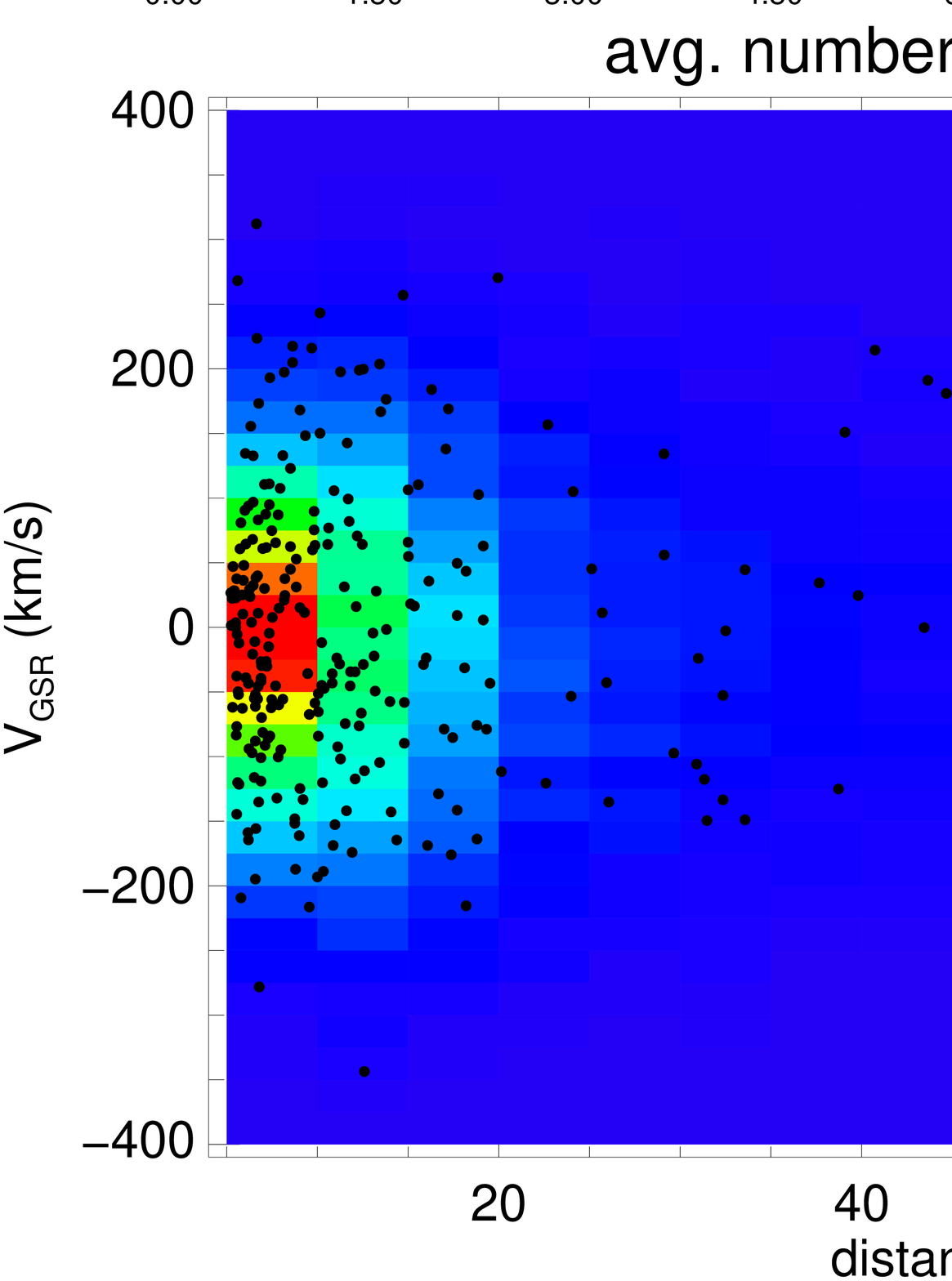}
\includegraphics[width=0.33\textwidth, trim=0.3in 0.3in 0.5in 0.3in, clip]{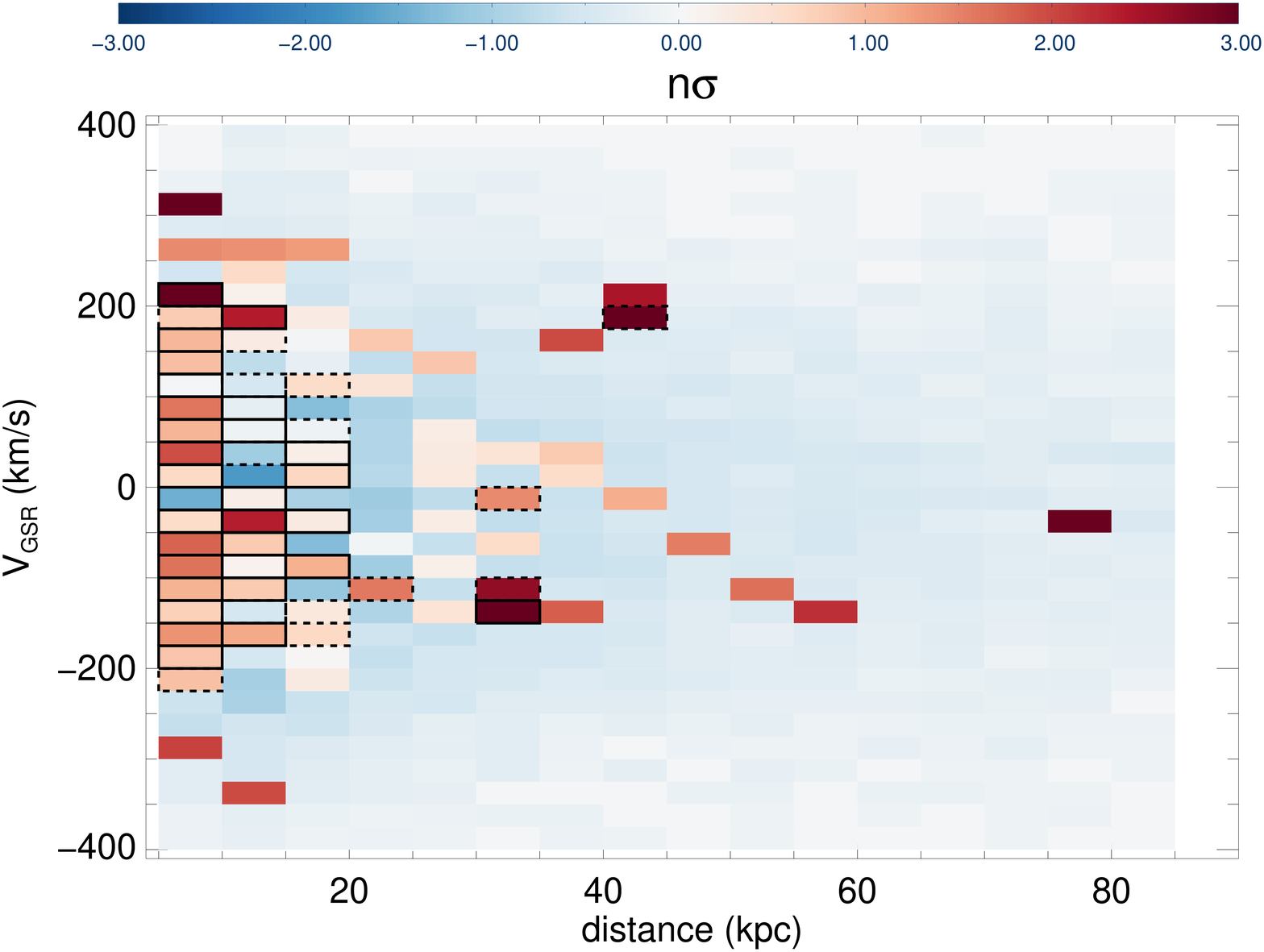}
\caption{Two examples of our search for SHARDS in $12^\circ \times 12^\circ$ fields. In each row, the left panel shows the histogram of (Galactocentric) line-of-sight velocities ($V_{\rm GSR}$) measured by LAMOST. The dashed line represents the average (with gray regions showing the $2\sigma$ deviation) expectation from 100 Galaxia models in the same sky area, observed with the LAMOST selection function. The middle panels are 2D histograms in $V_{\rm GSR}$ and distance, with the color scale encoding the average number of stars in each bin from the 100 Galaxia realizations, and black points representing the LAMOST observations. The right panels color-code the deviation (in number of $\sigma$ deviations) of LAMOST data from model predictions in each bin. Red-orange colors represent excesses, and blue-colored bins are where LAMOST has fewer stars than expected. Bins that are outlined with a solid black line in the right-most column contain three or more stars, while those with a dashed outline have two stars. SHARDS are those bins with $>3\sigma$ excesses and $\geq2$ stars. The top row is for Field 74 at (RA, Dec) = (162.0$^\circ, 20.0^\circ$), and the bottom row is Field 78: (RA, Dec) = (210.0$^\circ, 20.0^\circ$). Field 74 has a prominent structure in velocity and distance due to the Sgr tidal stream. While no structure is obvious in the $V_{\rm GSR}$ histogram for Field 78, our algorithm identifies substructure in this field (bottom right panel). }
\label{fig:field74_78_rv_dist}
\end{figure*}

We initially considered a method similar to that of \citet{sra+09},
which treated each SDSS/SEGUE plate individually. However, because the
LAMOST limiting magnitude is fairly bright ($r \sim 17.8$), our study
is limited to giant stars, of which there are very few on each
individual LAMOST line of sight (referred to as
``plates''). Nevertheless, when the entire $> 3$~million-star data set
currently available from LAMOST is considered, we have a sufficient
number of giants to statistically probe the halo. Freeing ourselves of
the constraints of individual plate areas helps us to probe larger
spatial scales than were available in the ECHOS searches, but the
varying selection function from plate to plate must then be accounted
for in our work, as discussed further in Sections 3 and 4 below.

Our method of searching for substructures proceeds as follows. We
divide the sky into $12^\circ \times 12^\circ$ regions in right
ascension (RA) and declination (Dec), covering the entire northern sky
visible to LAMOST ($-10^\circ <$~Dec~$< 60^\circ$). This scale was
chosen to provide a large statistical sample in each region, while not
being so large that the Galactic stellar populations vary dramatically
across the region.  In order to identify substructure in velocity and
distance, we need to know what the underlying (``smooth'') halo
populations (if there are any) look like. In other words, we wish to
assess what structures exist in {\it excess} over what we expect from
the smooth halo distributions. We adopt the Galaxia \citep{sbj+11}
model for this purpose. Galaxia generates expected observational
catalogs by sampling from empirically-based density and velocity
prescriptions for Milky Way stellar populations. In particular,
Galaxia uses an oblate power-law halo density profile (following
\citealt{rrd+03}) with a velocity ellipsoid of $(\sigma_r,
\sigma_\theta, \sigma_\phi) = (141, 75, 75)$~km~s$^{-1}$. Each
instance of the model produces a predicted set of observations given
prescribed color/magnitude ranges and a region of sky. Sampling from
the underlying distributions begins with a random number seed, which
can be changed to generate distinct model observations in a given sky
area. Thus, to sample the smooth distributions, we choose to run
multiple instances of the Galaxia model initialized with different
random number seeds, and build a set of mock catalogs with which to
statistically compare the LAMOST observations.

We stress that our comparisons throughout this work tell us how much
substructure is present {\it relative to the Galaxia model predictions
for the underlying halo}, so that some fraction of any observed
discrepancies may be due to deficiencies in the Galactic components
encoded in the model.

Finally, we must account for the inhomogeneous way that the LAMOST
survey targets are selected. Plates in the LAMOST survey are separated
by magnitude ranges into ``Very Bright'' (VB; $r < 14$), ``Bright''
(B; $14 < r < 16.8$), ``Medium'' (M; $16.8 < r < 17.8$), and ``Faint''
(F; $r > 17.8$) plates so that a variety of observing conditions can
be accommodated, while also limiting the magnitude range included in
each plate to ensure that sufficient $S/N$ is achieved for a large
fraction of stars in each observation. This means that the magnitude
range probed in a given region of the sky may differ dramatically from
a neighboring region. In such cases, using only the distance and
velocity distributions of observed stars without accounting for the
selection function of stars that were targeted may strongly bias the
results. We thus need to ``observe'' the mock catalogs using the
LAMOST selection function in order to obtain a valid comparison.

The process of searching for excess stellar substructures proceeds as follows:

\begin{enumerate}

  \item Run 100 simulations (each initialized with different random seeds) for that field with Galaxia (including only thick-disk and halo components, and using the same constraints that were applied to isolate halo stars from the LAMOST data; see Section~\ref{sec:data}).
  \item In each simulation, randomly select a star that is within 0.025 mags in $K_{\rm S}$ vs. $(J-K_{\rm S})$ color-magnitude space (or, if the nearest star is more than 0.025 mags away, try 0.05 mags, then if still none, simply select the nearest) of each LAMOST object in the Galaxia output. In this way, we ``observe'' the Galaxia model in the same way that LAMOST has observed that part of the sky. This results in a sample consisting of 92\% mock halo stars and 8\% thick-disk stars (averaged over all of the mock observations, and over the entire sky), as tagged by Galaxia.
  \item Bin the simulated observations into 25 km~s$^{-1}$ by 5 kpc bins in the $V_{\rm GSR}$ vs. distance plane, then calculate the average number of stars found in each of these bins over the 100 simulated observations of that field of view. We take the standard deviation over the 100 mock catalogs as an error bar on the counts in each bin.
  \item Bin the observed LAMOST data in the same way, and calculate the residual number of stars for each bin, expressed as a deviation in the sense $n_\sigma = (N_{\rm LAMOST} - N_{\rm model})/\sigma_{\rm model}$, where $N_{\rm LAMOST}$ is the number of observed stars in each bin, $N_{\rm model}$ is the average number predicted by the model, and $\sigma_{\rm model}$ is the standard deviation of the model counts over the 100 simulations.
  \item Define an ``excess'' as a bin that has $>3\sigma$ excess of observed stars. Rather than trying to identify which stars should be removed as ``background'' stars, we retain all stars from each bin in which we have identified an excess. Because we expect some excesses simply due to statistical fluctuations (and we are not considering deficient bins where there are fewer stars than predicted), this method includes more stars than are actually present in SHARDs. We explore this effect further in Section~\ref{sec:smooth_compare}, and assess the fraction of ``smooth halo'' stars that are identified as ``excess'' by our method. For the analysis in Sections 4 and 5, we subtract the expected number of background stars to derive the net number of excess stars in SHARDS.

\end{enumerate}

Figure~\ref{fig:field74_78_rv_dist} shows examples of two fields from
our study. The upper row is field 74, centered at (RA, Dec) =
(162.0$^\circ, 20.0^\circ$), and the lower row is field 78, at (RA,
Dec) = (210.0$^\circ, 20.0^\circ$). The left panels in each row show
the observed line-of-sight velocity distribution (in a Galactocentric
frame), $V_{\rm GSR}$, from LAMOST as a solid black histogram. The
dashed line in each of these panels shows the mean velocity from the
100 Galaxia model realizations in that field, with the $2\sigma$
region given by the gray shaded region (where $\sigma$ is the standard
deviation of the model counts in each bin). These panels show a stark
difference in the amount of velocity substructure exceeding the
$>2\sigma$ level. Field~74 (the upper panel) shows an obvious excess
centered at $V_{\rm GSR} \sim -100$~km~s$^{-1}$ that exceeds the
$2\sigma$ shaded region in four consecutive bins. This field also has
a high-velocity peak that exceeds the expected velocity. In contrast,
field 78 (lower panels) has no velocity peaks exceeding the $2\sigma$
region in the $V_{\rm GSR}$ histogram.

We now show that with the addition of distance information to the
clustering search, substructures emerge even in cases such as field
78, which shows no obvious velocity structure in the lower-left panel
of Figure~\ref{fig:field74_78_rv_dist}. The center panels of both rows
in Figure~\ref{fig:field74_78_rv_dist} display 2D histograms with bins
of 25~km~s$^{-1}$ in $V_{\rm GSR}$ and 5~kpc in (heliocentric)
distance. The color of each bin corresponds to the average number of
stars in that bin from the 100 model realizations, and the black
points are the LAMOST data. Given this binning and the standard
deviation, $\sigma$, for each bin, one can simply compare the expected
number count to the observed count from LAMOST. The deviation from
model expectation, $n_\sigma = (N_{\rm LAMOST} - N_{\rm
model})/\sigma_{\rm model}$, is mapped in the right panels of
Figure~\ref{fig:field74_78_rv_dist}, where the red-orange colors
represent excesses relative to the model, and blue bins are
deficits. Comparing the 2D maps to the velocity histograms, it is
clear that the obvious velocity peak in field~74 (upper row) is also
coherent in distance, as expected for a tidal debris structure. This
feature is most likely related to the Sagittarius stream, and has
velocity and distance consistent with those from the model of
\citet{lm10a} and the SDSS observations of \citet{bke+14}. Note also
that some features that appear as strong excess bins in the right-most
panels contain only a single star; because the average occupancy of
many bins (given the model convolved with the LAMOST selection
function) is well below one star for many bins, a single star may be a
statistically significant excess. However, for subsequent analysis we require bins to have two or more stars for consideration as SHARDS. 

\begin{figure*}[!t]
\includegraphics[width=0.33\textwidth, trim=0.3in 0.3in 0.25in 0.3in, clip]{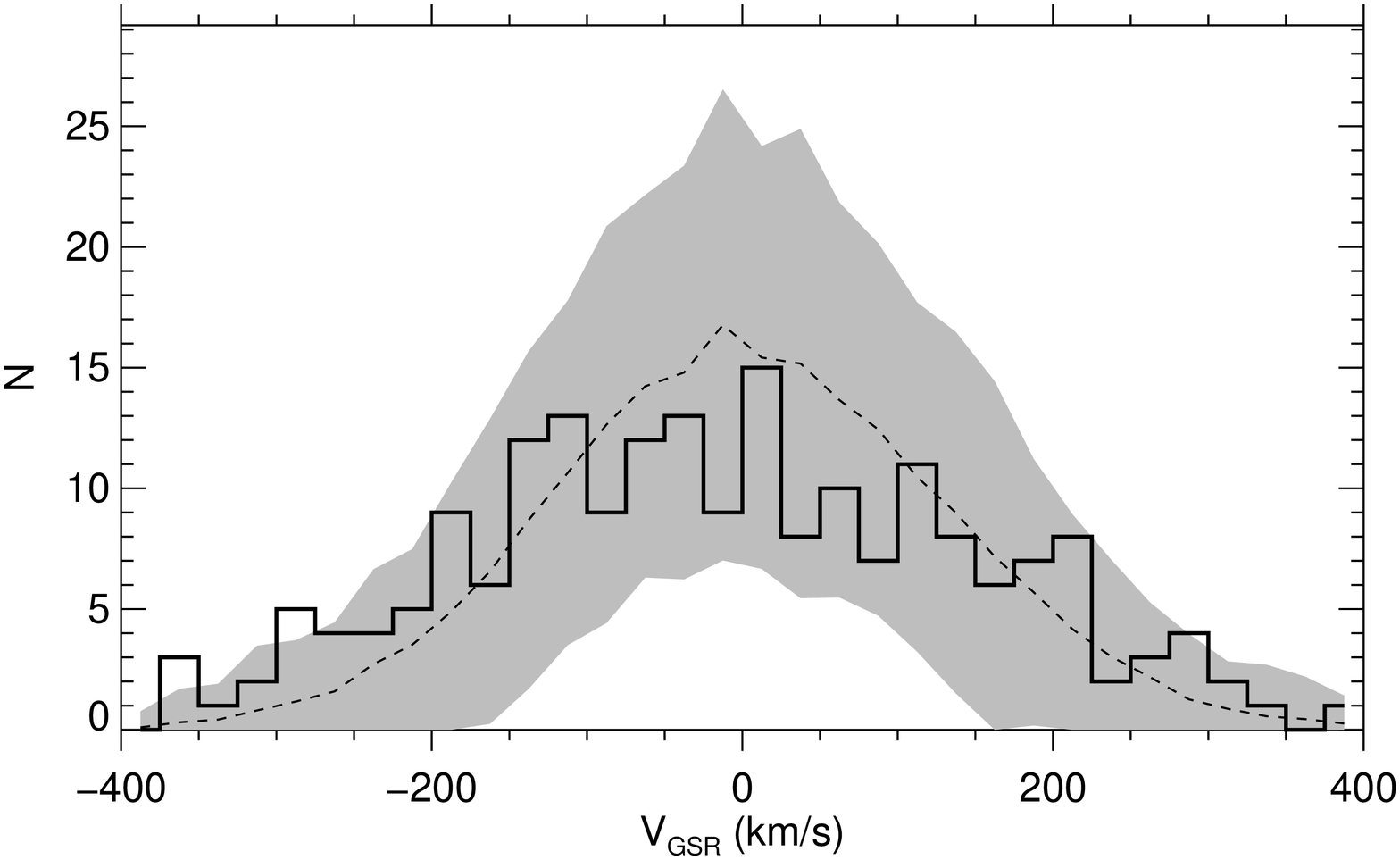}
\includegraphics[width=0.33\textwidth, trim=0.3in 0.3in 0.5in 0.3in, clip]{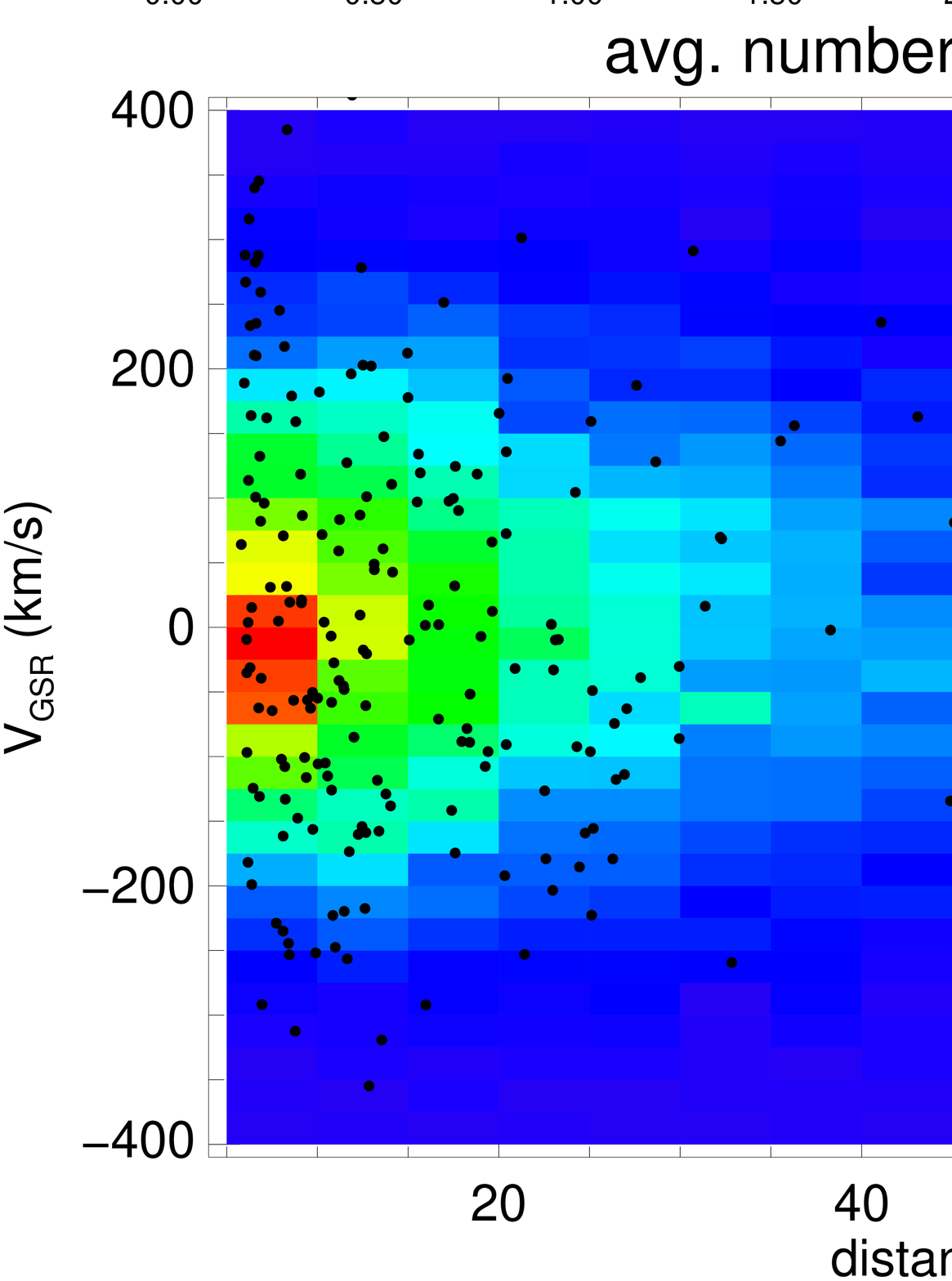}
\includegraphics[width=0.33\textwidth, trim=0.3in 0.3in 0.5in 0.3in, clip]{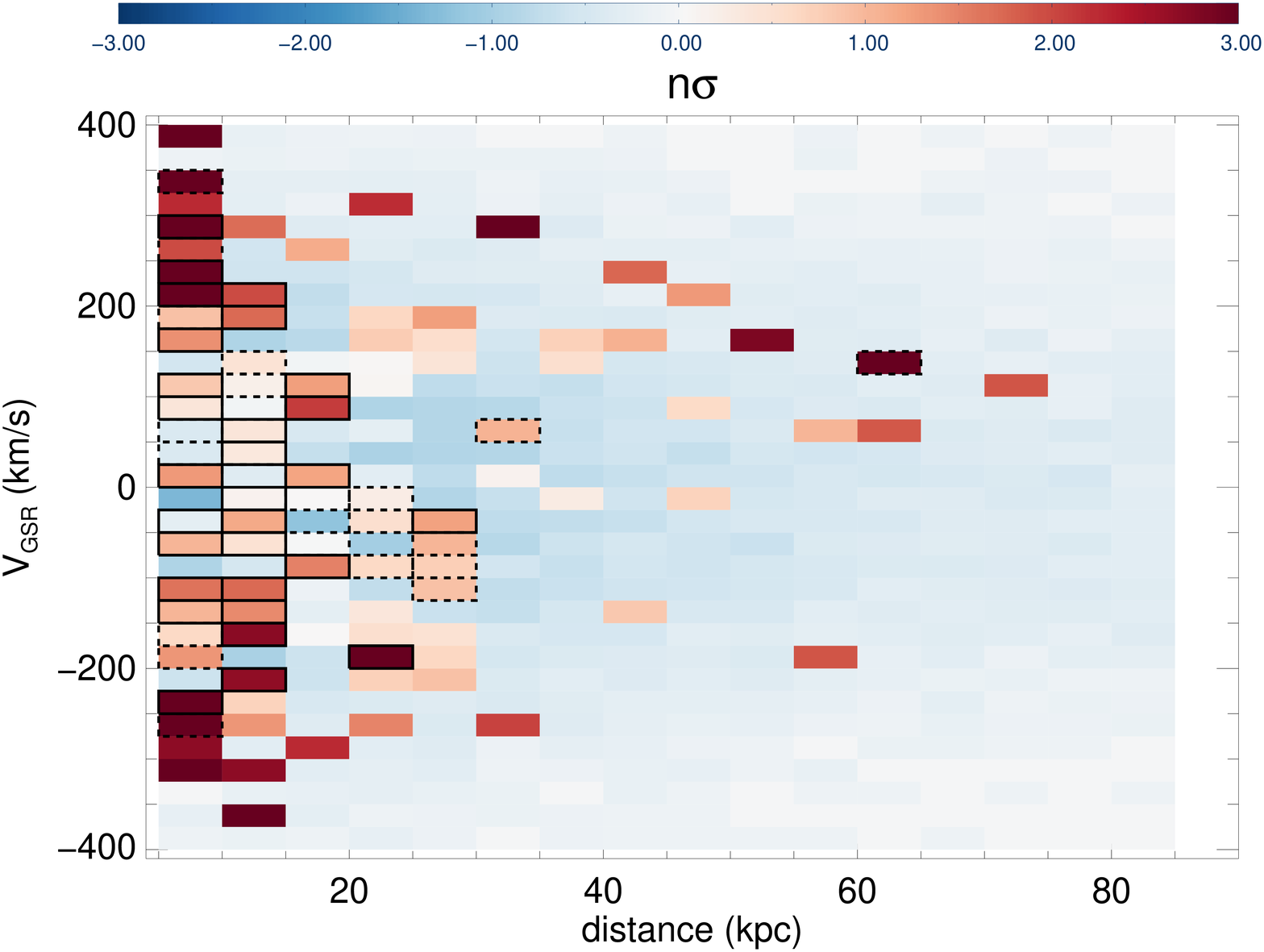}\\
\includegraphics[width=0.33\textwidth, trim=0.3in 0.3in 0.25in 0.3in, clip]{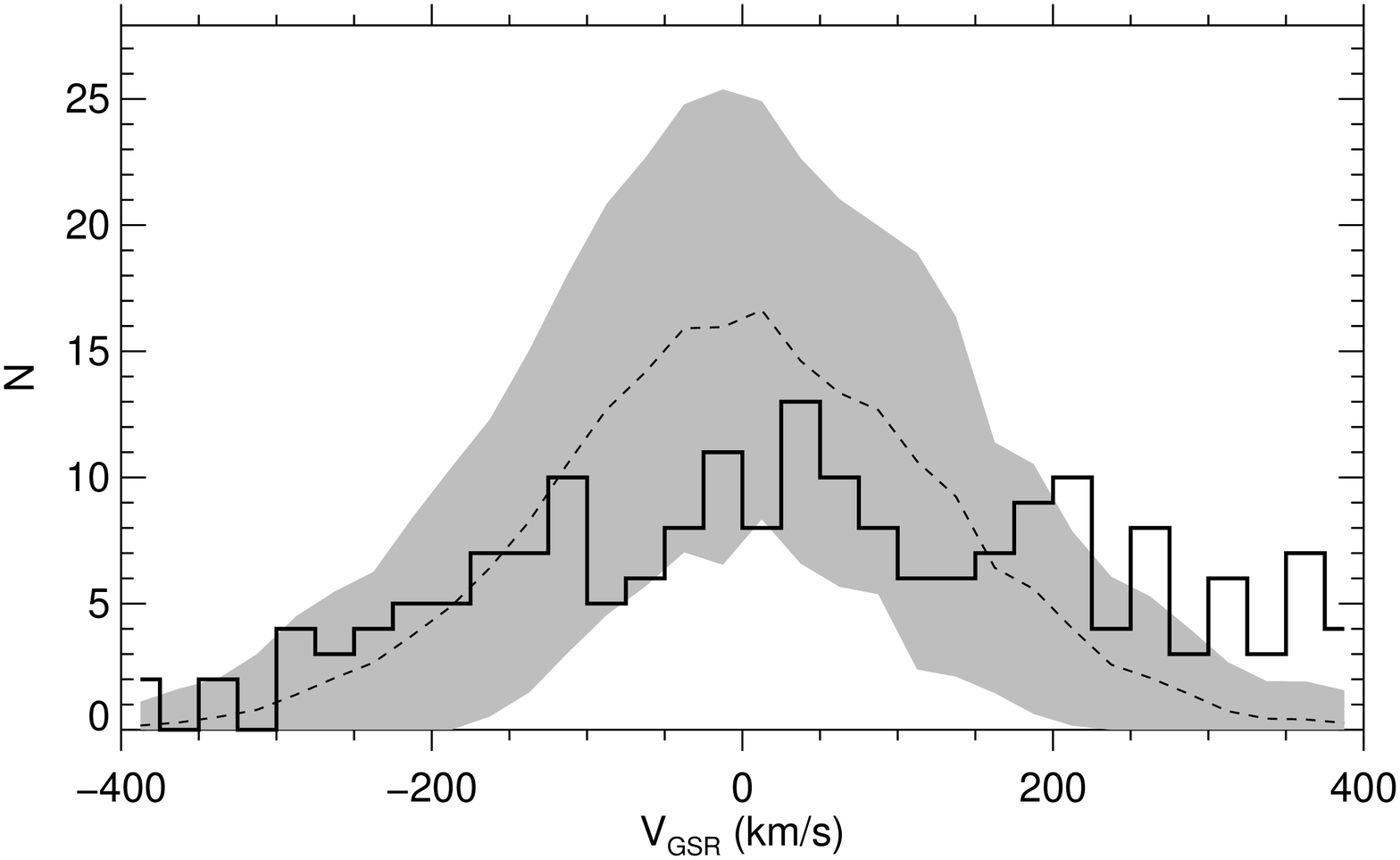}
\includegraphics[width=0.33\textwidth, trim=0.3in 0.3in 0.5in 0.3in, clip]{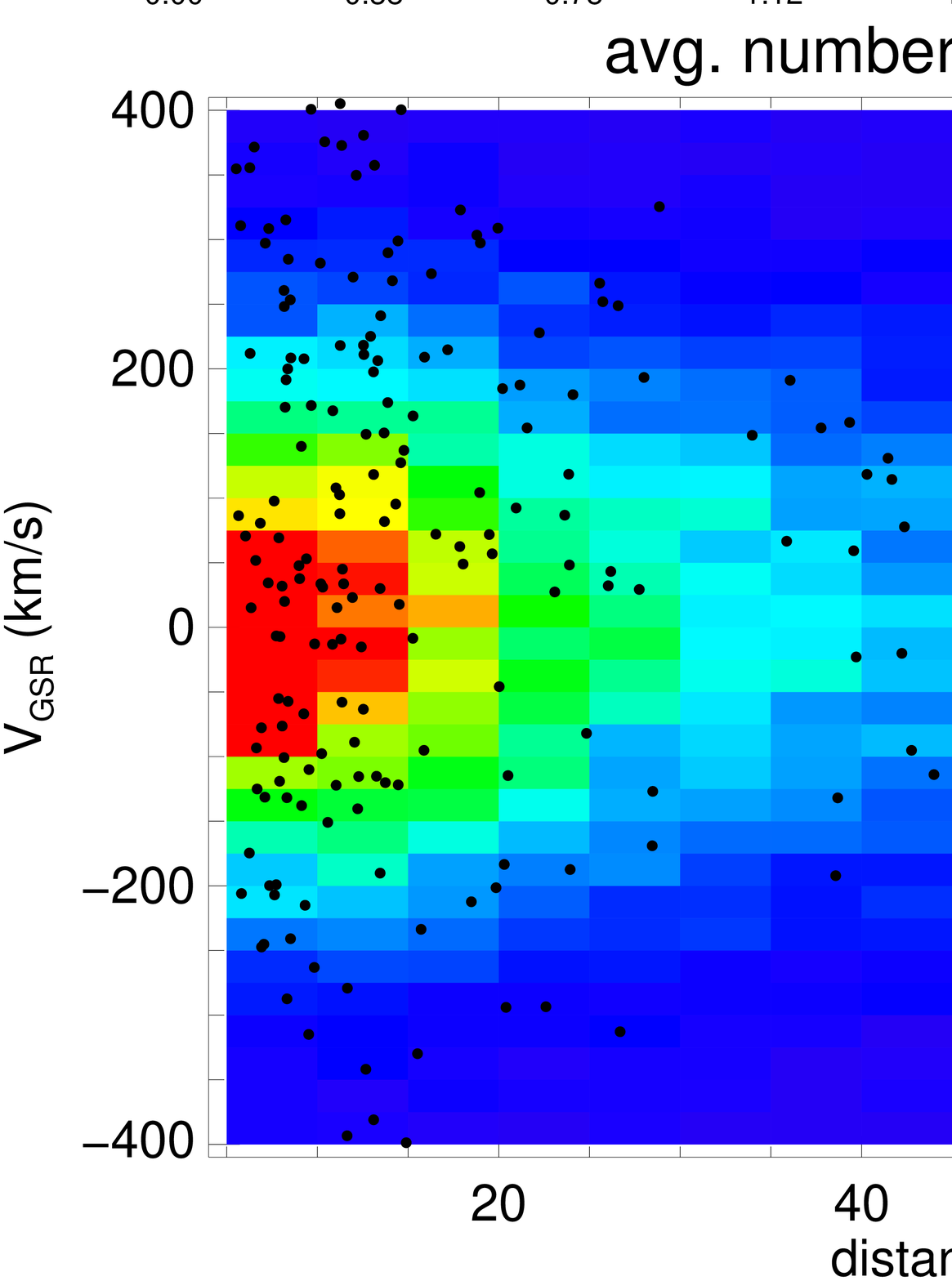}
\includegraphics[width=0.33\textwidth, trim=0.3in 0.3in 0.5in 0.3in, clip]{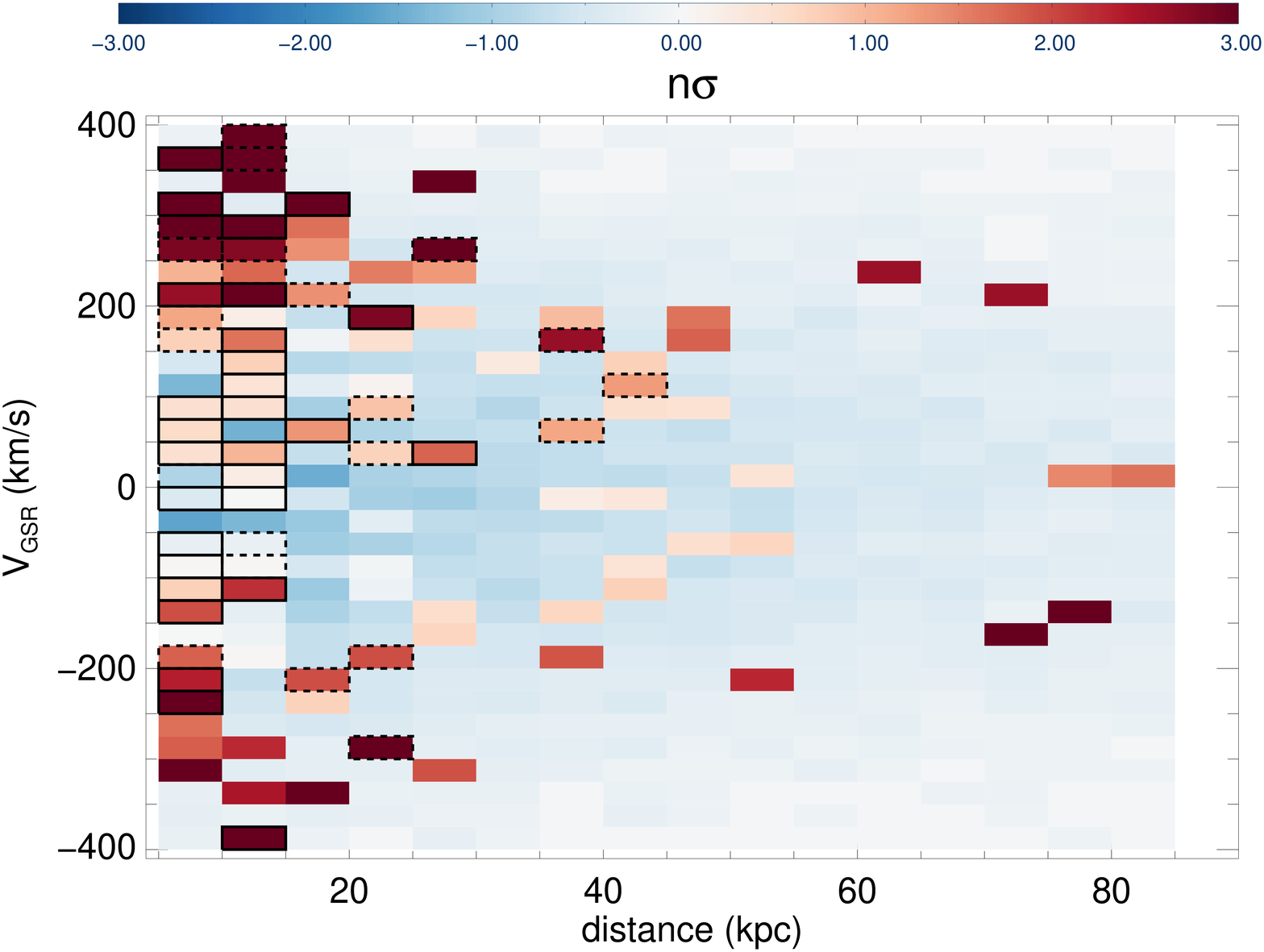}
\caption{As in Figure~\ref{fig:field74_78_rv_dist}, but for Bullock \& Johnston models 1 and 8 along the line of sight of Field 74: (RA, Dec) = (162.0$^\circ, 20.0^\circ$). Compare to the upper row of Figure~\ref{fig:field74_78_rv_dist}.}
\label{fig:BJ05_field74_rv_dist}
\end{figure*}

While field~78 (lower panels in Figure~\ref{fig:field74_78_rv_dist})
shows no obvious velocity structure in the left-most panel, there are
indeed significant excesses in velocity-distance phase space. We
define SHARDS as all bins containing at least two stars and greater than $3\sigma$ excess in
position-velocity phase space (i.e., bins in the right column of
Figure~\ref{fig:field74_78_rv_dist} that are above $3\sigma$), and
select all observed stars from these bins (note that in Figure~\ref{fig:field74_78_rv_dist}, bins with 3 or more stars are outlined with a thick solid line, and those with two stars with a dashed outline). In field~74, which has
Sagittarius debris obviously contributing significant substructure, 68
stars out of a total of 202 halo giants reside in SHARDS (after subtracting the background contribution in each bin). Field~78,
which shows no obvious velocity structure in the histogram,
nevertheless contains 17 stars in SHARDS out of a total of 263 in the
field. This illustrates the value of using additional dimensions of
information to seek substructures.

We divide the sky into 180 regions that are 12$^\circ \times 12^\circ$
in RA and Dec, spaced 12 degrees apart on the sky (with centers at
Declinations of $-4^\circ, 8^\circ, 20^\circ, 32^\circ, 44^\circ,$ and
$56^\circ$). Of these, there are 112 regions that have more than 10
LAMOST halo giants, the threshold we applied for inclusion in our
search, with a total of 10,481 stars. Then, we perform the same search
for SHARDS in each of these regions as in the fields illustrated
above. This yields a total of 1,140 (2,065) stars in $>3\sigma$
($>2\sigma$) excesses (after subtracting the model-predicted
``expected'' number), or $\sim10.9\%$ ($\sim19.7\%$) of the LAMOST
halo giants identified with SHARDS.

\subsection{Caveats}\label{sec:caveats}

The above result should not be directly interpreted as saying that
 $\sim20\%$ of the halo is in substructure. Before we can assess the
meaning of the absolute numbers of stars our method identifies in
SHARDS, we must consider the effects of the LAMOST selection function
and the expected ``false positives'' that the inhomogeneous selection
function would identify even if sampling a smooth halo. Additionally,
we test our method on halos created entirely from accreted satellites
in cosmologically-motivated simulations to assess the expected
signature of a purely accreted Milky Way halo.

We reiterate that our method identifies excesses above what is predicted by the smooth
halo prescription in Galaxia. Thus, any interpretation should
bear in mind that uncertainty about the shape, density profile, and
velocity distribution of the halo will cause discrepancies between the
model and observations. However, any one of these effects alone will
affect only one dimension of our 4D search for SHARDS, and thus would
be unlikely to induce clustered excesses. For example, if the density
profile used in Galaxia is too shallow, we would find excess stars in
some radial range, but it would be unlikely for their velocities to
also be similar. In addition, our requirement that each of the SHARDS have
more than a single excess star reduces the likelihood that model
inadequacies or small number samples would contribute to the detected
excesses.

Finally, we note that binning in $V_{\rm GSR}$ vs. distance likely
biases our results somewhat. For example, this technique may not
identify SHARDS that are split across bin boundaries, and thus contain
smaller numbers of stars in each separate bin. However, the binning is
valuable in allowing us to calculate statistics from the 100 model
realizations, so we deem the small biases to be acceptable. In fact,
most of the biases one can imagine in our rather straightforward
method are mitigated by comparing to models of smooth and
purely-accreted halos in an identical way to our treatment of LAMOST
data, as described below.

\begin{figure*}[!t]
\includegraphics[width=0.33\textwidth, trim=0.3in 0.3in 0.25in 0.3in, clip]{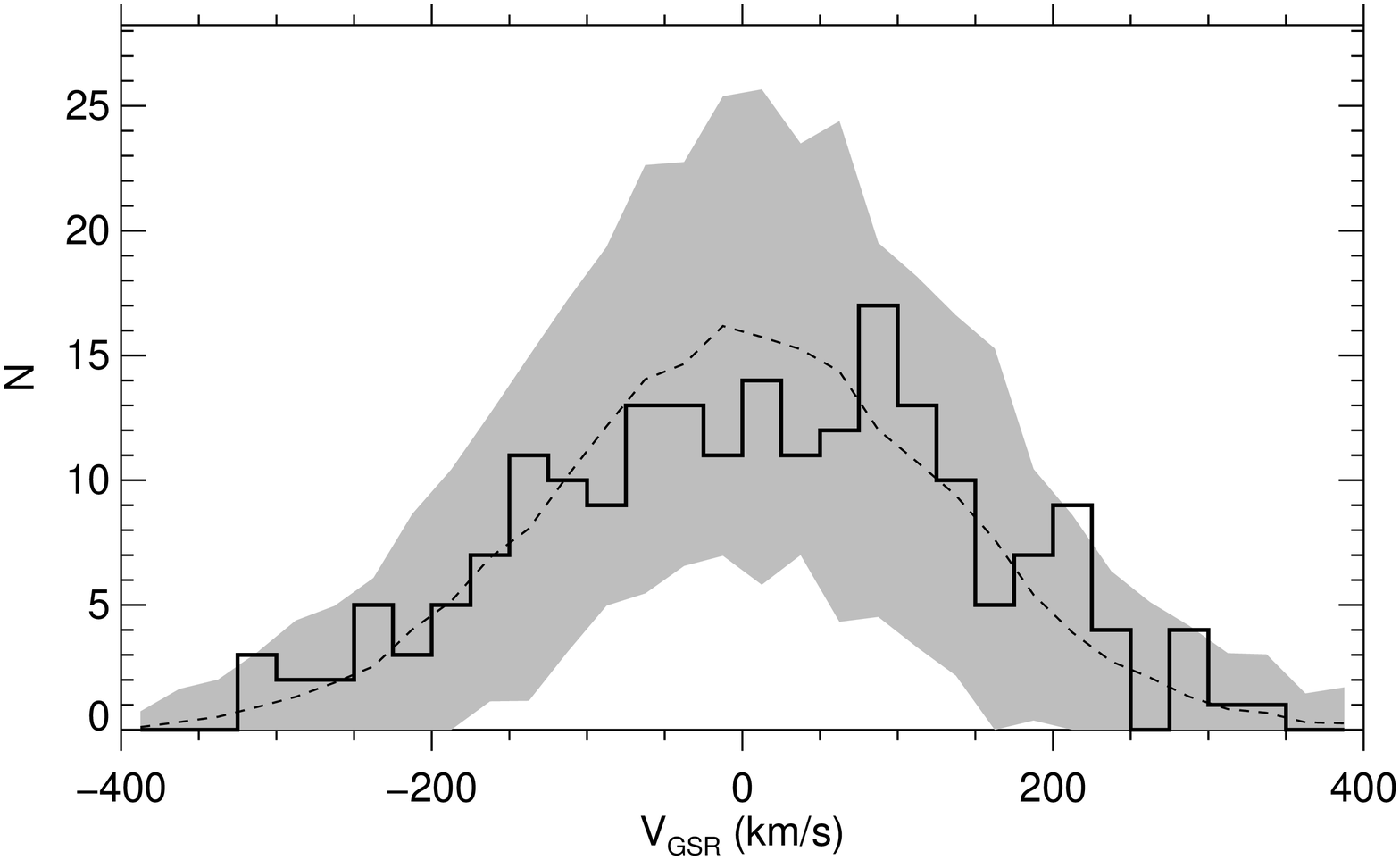}
\includegraphics[width=0.33\textwidth, trim=0.3in 0.3in 0.5in 0.3in, clip]{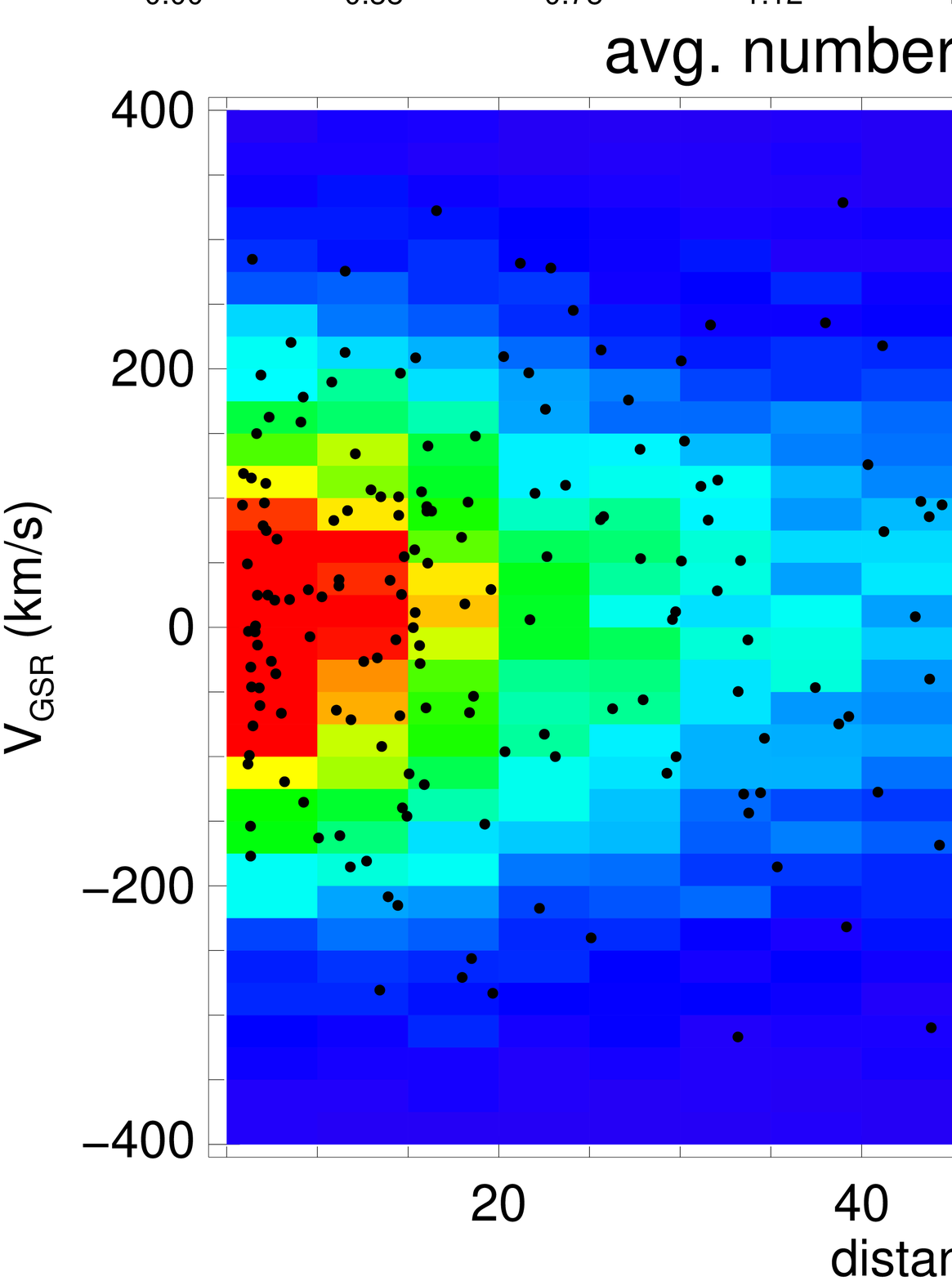}
\includegraphics[width=0.33\textwidth, trim=0.3in 0.3in 0.5in 0.3in, clip]{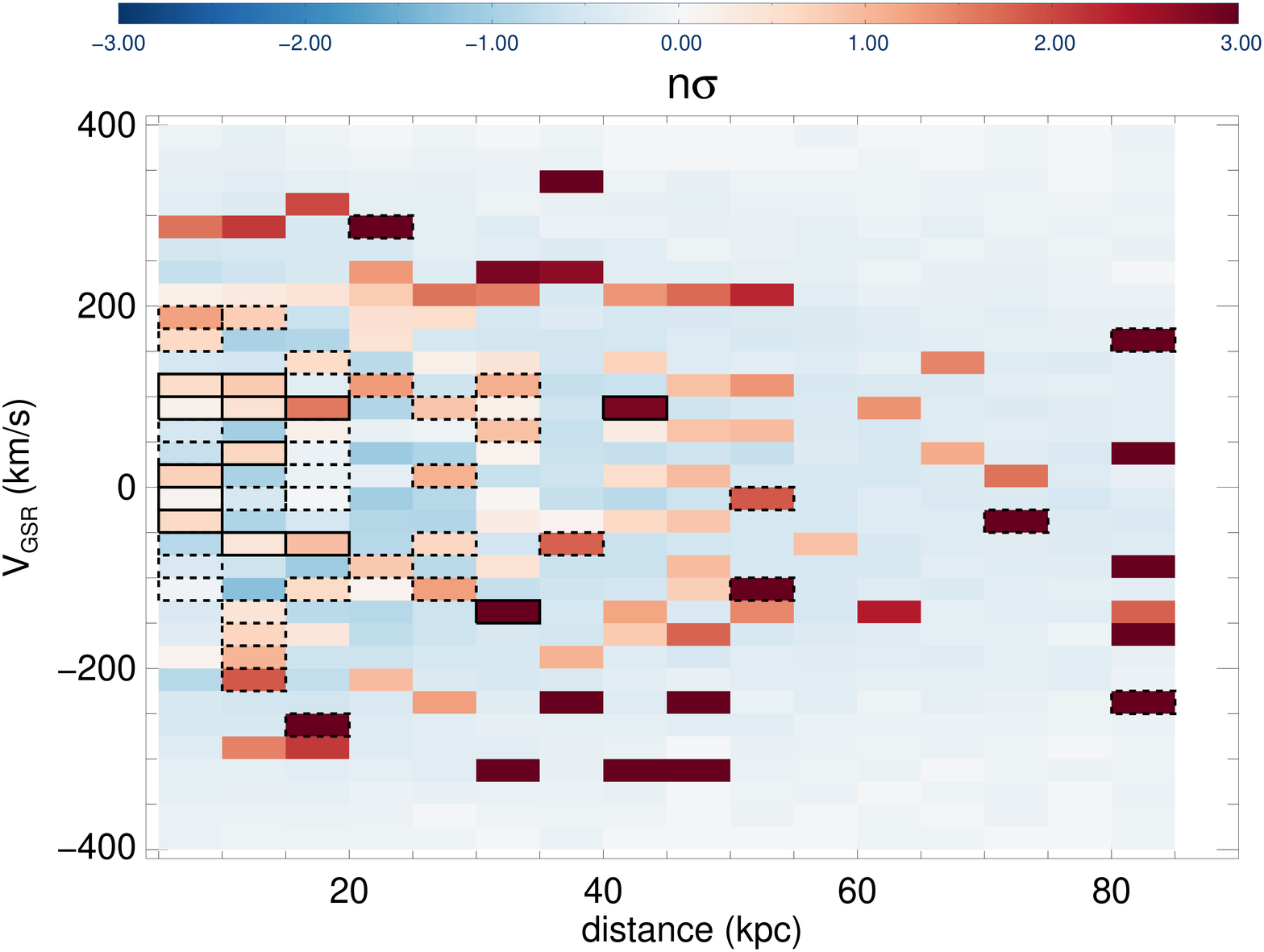}
\caption{As in Figure~\ref{fig:field74_78_rv_dist}, but for Galaxia smooth halo models for Field 74: (RA, Dec) = (162.0$^\circ, 20.0^\circ$). Compare to the upper row of Figure~\ref{fig:field74_78_rv_dist} and to Figure~\ref{fig:BJ05_field74_rv_dist}.}
\label{fig:smooth_field74_rv_dist}
\end{figure*}

An additional important effect is our choice of a fixed 5~kpc bin width in distance. The typical uncertainties on individual stellar distances may be as high as $\sim30\%$ \citep{cln+15}, so that the distance errors are larger than the bin size for distances $\gtrsim 17$~kpc. With a fixed bin size of 5~kpc, the distance uncertainties will have the effect of scattering stars between bins, thus spreading the signal of any SHARDS over multiple bins in distance (note that the 25~km~s$^{-1}$ velocity bins are larger than the typical velocity error of 5-10~km~s$^{-1}$, so this is not an issue in the RV dimension). The net effect of this will be that some legitimate SHARDS are not identified as such, since a bin that should have been an excess will be spread out over multiple bins, with the number of stars in each bin being below the threshold for identification as a SHARDS. [Note that this also means that the detected SHARDS may actually contain more stars than identified by our technique, with some members missed because they scattered into adjacent bins.] We have tested the magnitude of this effect by using our SHARDS algorithm with variable-width bins spanning distance ranges of 5-10~kpc, 10-20~kpc, 20-40~kpc, and 40-80~kpc. These larger bins yield similar total numbers of LAMOST halo giants in SHARDS as were found with fixed bins, with 1,057 (1,885) stars, or $\sim10.1\%$ ($\sim18.0\%$), in $>3\sigma$ ($>2\sigma$) excesses (after subtracting the model-predicted
``expected'' number). We thus conclude that the fixed distance bin size has not significantly biased our results; this will be explored in more detail in future contributions.

\section{Model Comparisons}\label{sec:compare}

To place our observations into context, we first examine the results
from running our code on the model halos from
\citet[BJ05]{bj05}. These halos are built purely from accreted
satellites generated from cosmologically-based merger trees and
accretion histories. We thus use our mock observations of these halos
as a test case to see what one might expect to observe if the Galactic
halo consists entirely of satellite debris.

Another test we use to place our observations in context is to
generate model halos using only the smooth halo and thick disk
prescriptions in Galaxia, then observe these halos in the same way
that LAMOST samples the sky. This provides a picture of how many
statistical fluctuations and chance groupings are detected as SHARDS
by our technique in the case of a smooth halo.

\subsection{Comparison to Bullock \& Johnston ``pure accretion'' halo}\label{sec:bj_compare}

Even if the Galactic halo is made up entirely of accreted satellites,
the debris from ancient accretion events may be phase mixed and thus
no longer visible as substructure in phase space. Given that most of
the accreted satellites were assimilated long ago, the fraction of
halo stars that is in observable substructure will not be
100\%. Furthermore, the fraction that is in substructure at present
depends on the specific merger history of the Galaxy. To assess what
we expect to see in the case that the Galactic halo is made up wholly
of satellite remnants, we use the model halos from BJ05. In this work,
cosmologically-motivated initial conditions were used to give the
properties of accreted satellites, each of which was then modeled via
$N$-body simulations. The results of all of these $N$-body models for
a given host galaxy were compiled together to create a simulated
stellar halo created from the simulated, accreted satellites. The
satellite properties in BJ05 match those of known Milky Way dwarfs,
and the resultant halos have roughly the same total luminosity as the
Milky Way. In these models, the destroyed satellites that contribute
most of the stellar halo are accreted early (at least $\sim9$~Gyr
ago), and their debris is mostly smoothly distributed and located in
the inner halo (inside $R_{\rm GC} \sim 10$~kpc). Most visible
substructure is seen in the outer ($R_{\rm GC} > 20$~kpc) halo, where
more recent accretion events deposit their debris.

The 11 simulated halos from BJ05 have been implemented in Galaxia, so
that we can take simulated observations of these models and then
search for SHARDS in them using the same algorithm we have used for
the LAMOST data. For each of the 11 simulations, we generate a catalog
using Galaxia, then observe this model by selecting the star that is
closest to each star from LAMOST in $K_0$ vs. $(J-K)_0$. This mock
observed catalog is then passed through the routines outlined above to
detect all SHARDS in the BJ05 model (given the LAMOST selection
function). An example of the results for Field 74 from BJ05 models 1
and 8 (note that these two halos were selected arbitrarily) is given
in Figure~\ref{fig:BJ05_field74_rv_dist}. These mock observational
results can be compared directly with those from the top row of
Figure~\ref{fig:field74_78_rv_dist}, which shows the LAMOST results
for Field 74. The velocity distribution (left panels of
Figure~\ref{fig:BJ05_field74_rv_dist}) of the BJ05 halos is clearly
broader than the expectations from Galaxia, with more stars in the
wings of the $V_{\rm GSR}$ histograms and a deficit (relative to the
Milky Way halo prediction) near $V_{\rm GSR} \sim 0$~km~s$^{-1}$. In
the middle and right-hand panels of
Figure~\ref{fig:BJ05_field74_rv_dist}, it becomes clear that most of
the high-velocity excess in the mock halos is nearby, well-mixed
debris, rather than distant distinct substructures. One obvious
difference between these mock halos and the LAMOST observations in
Field~74 is the dominance of the coherent Sagittarius debris structure
in the LAMOST data; no such obvious substructures are visible in the
BJ05 models. There is a feature in halo 8 (lower panels) that
stretches from roughly ($V_{\rm GSR}$, dist) $\sim (250$~km~s$^{-1}$,
25~kpc) to ($V_{\rm GSR}$, dist) $\sim (100$~km~s$^{-1}$, 45~kpc),
similar to the caustic-like features expected from some accretion
relics (e.g., \citealt{jbs+08, sh13}). However, in general these mock
halos show low levels of substructure at all radii, as might be
expected from a melange of accretion debris. Of 202 halo giants in
Field~74, there are 13 and 31 stars identified in SHARDS in halos 1
and 8 from BJ05, respectively. In the LAMOST observations, this number
is 68 of 202; however, $\sim30-35$ of these are likely related to the
Sgr stream, so that the total is $\sim35$ of non-Sgr stars in SHARDS
in this field. Thus the pure-accretion halo predictions are consistent with the number seen in LAMOST if we exclude Sgr, but predict fewer
stars in SHARDS than observed when Sgr debris is included. In fact, in only one (halo 7, with 79) of the field 74 regions of the BJ05 halos do we find as many stars in SHARDS as we have observed with LAMOST.

\begin{figure*}[!t]
\includegraphics[width=0.5\textwidth]{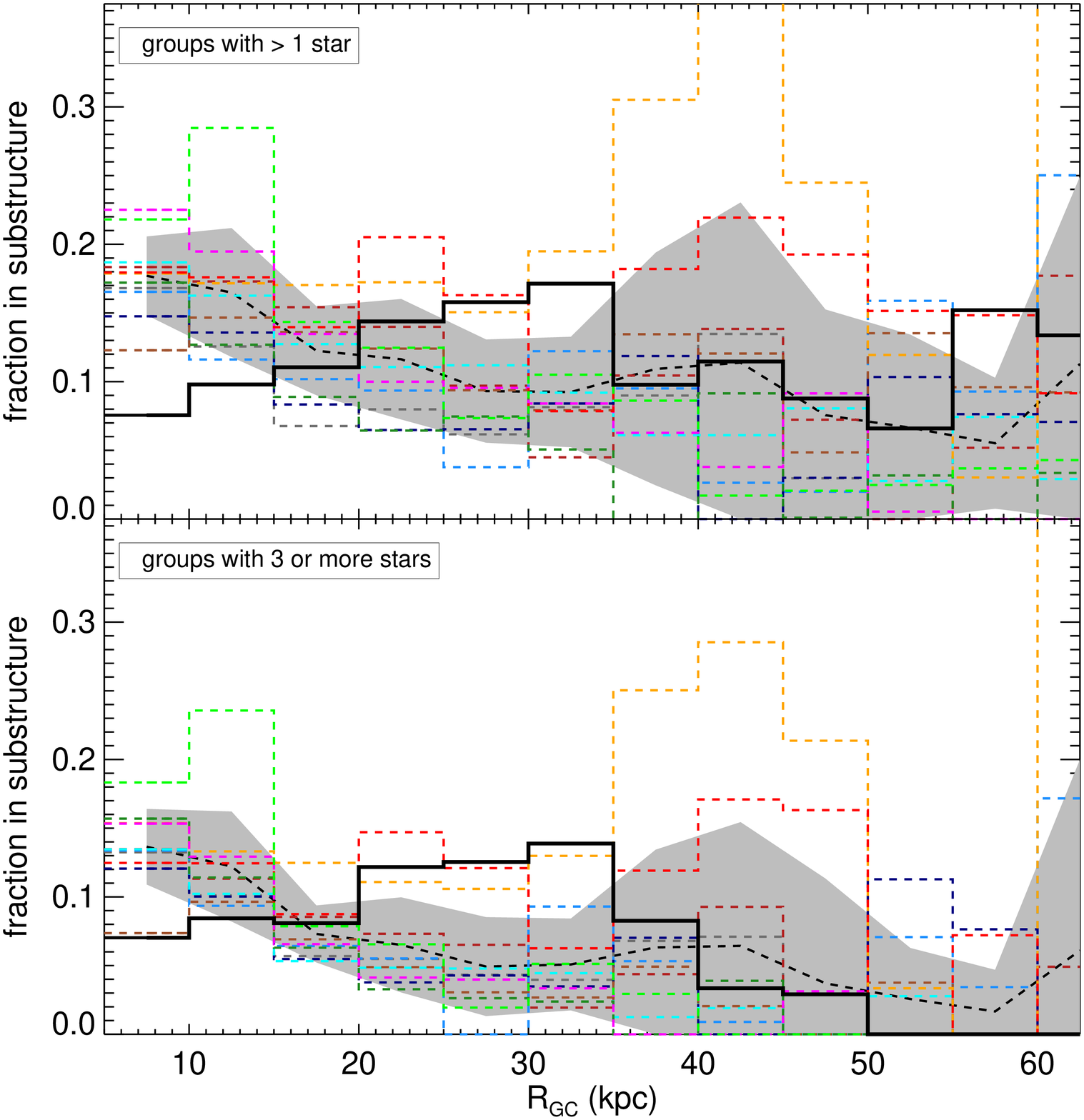}
\includegraphics[width=0.5\textwidth]{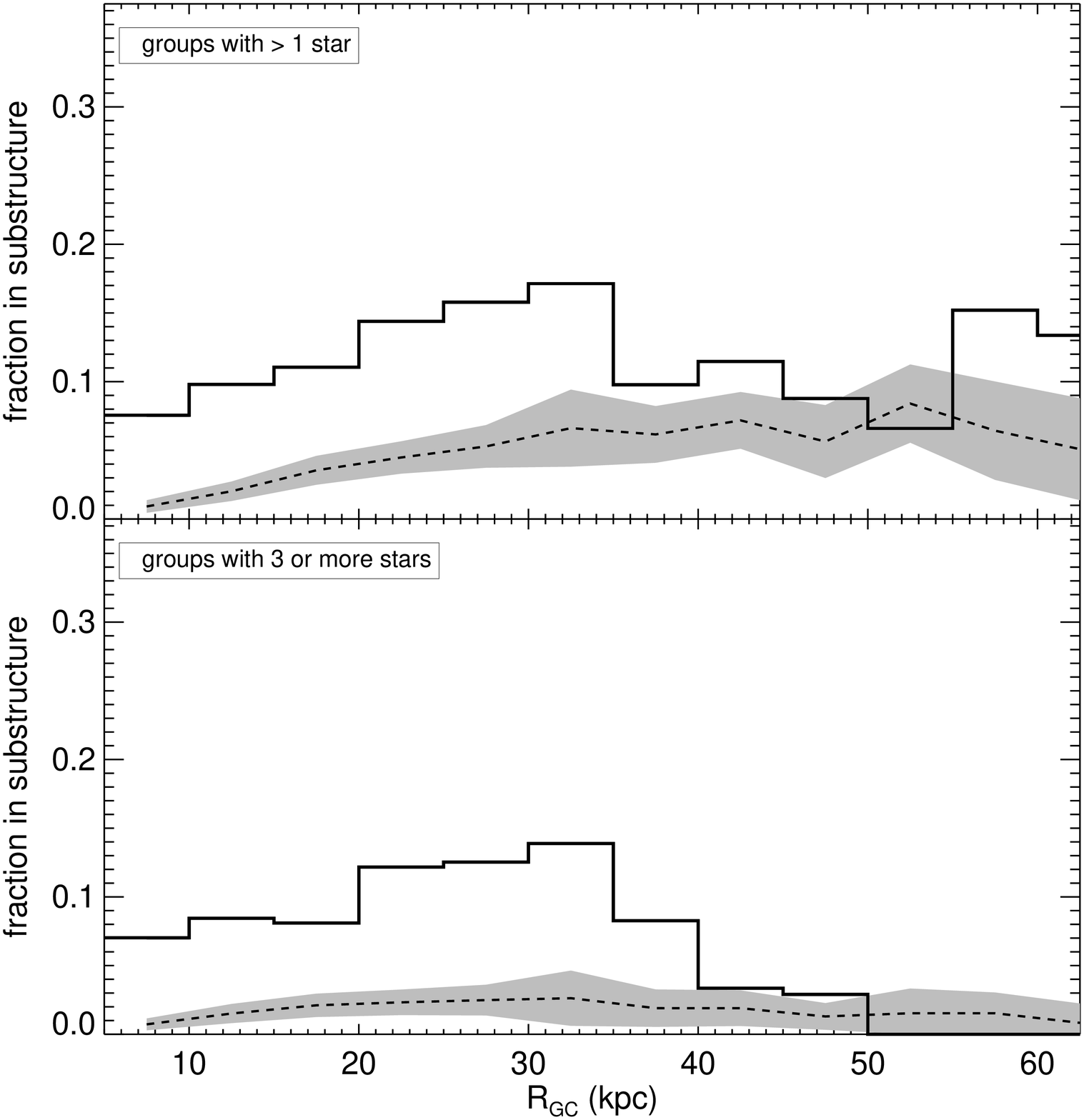}
\caption{Each panel shows the fraction of observed stars, after subtracting the expected number in each bin, in LAMOST DR1-3 that are identified as SHARDS in each 5 kpc bin in Galactocentric radius, $R_{\rm GC}$ (solid black histogram). The upper row gives the fraction of all stars identified as $> 3\sigma$ excesses in bins having more than one star, and the bottom panels require 3 or more stars per group. Gray shaded regions are the average and $1\sigma$ variation from mock observations of all 11 of the \citet{bj05} halos (left panels), and the average of 10 Galaxia simulations of the smooth halo (right panels). Colored lines in the left column are the results for each of the 11 model halos; these illustrate the variety of accretion histories present in the models.}
\label{fig:substruct_frac}
\end{figure*}

\subsection{Comparison to a ``smooth'' model halo}\label{sec:smooth_compare}

Given the sparse selection of halo giants from LAMOST, it is likely
that some stars will be identified as SHARDS simply due to statistical
noise from sampling the smooth halo distribution. In addition to the
inhomogeneous subsampling of the halo, uncertainties in the distances
(typically about 20-30\%, related mostly to stellar parameter errors;
see \citealt{cln+15}) can also shift stars from their true location in
the 2D plane of $V_{\rm GSR}$ vs. distance, making some stars look
like outliers compared to the expected populations. To quantify these
effects, we generate a model halo using only the smooth prescriptions
for the halo and thick disk that are implemented in Galaxia. By
performing mock observations of these halos in the manner described in
Section~\ref{sec:bj_compare}, we generate a catalog of what would be
expected in the LAMOST halo-giant sample if the underlying
distributions are smooth.

We run this observed catalog of ``smooth halo'' stars through our
SHARDS pipeline. Plots of the results for field 74 are shown in
Figure~\ref{fig:smooth_field74_rv_dist}. The velocity histogram
reproduces the underlying distribution well, with only a handful of
bins making excursions larger than one sigma. However, our method does
find a number of stellar excesses in this field. We find that 18 of
202 stars in Field~74 are in SHARDS (compare to 68 of 202 in LAMOST,
or $\sim35$/202 if Sgr is excluded). We repeat this exercise for 10
different Galaxia smooth halo models generated with different random
seeds. The average number of stars in SHARDS is 11/202 (ranging from
5-18) in Field~74, while the average number in the 11 BJ05 halos in
this same field is 40/202 (range: 13-79). This gives an idea of the
level of ``background'' contributed by sampling the smooth halo with
the LAMOST selection function, as well as the simplistic method of
denoting all stars in bins with $>3\sigma$ excesses of two or more stars as SHARDS. The
smooth halo simulation and the BJ05 pure accretion halos predict a
range that helps us to interpret the number of LAMOST halo stars in
SHARDS according to our algorithm.

\begin{figure*}[!t]
\includegraphics[width=0.51\textwidth]{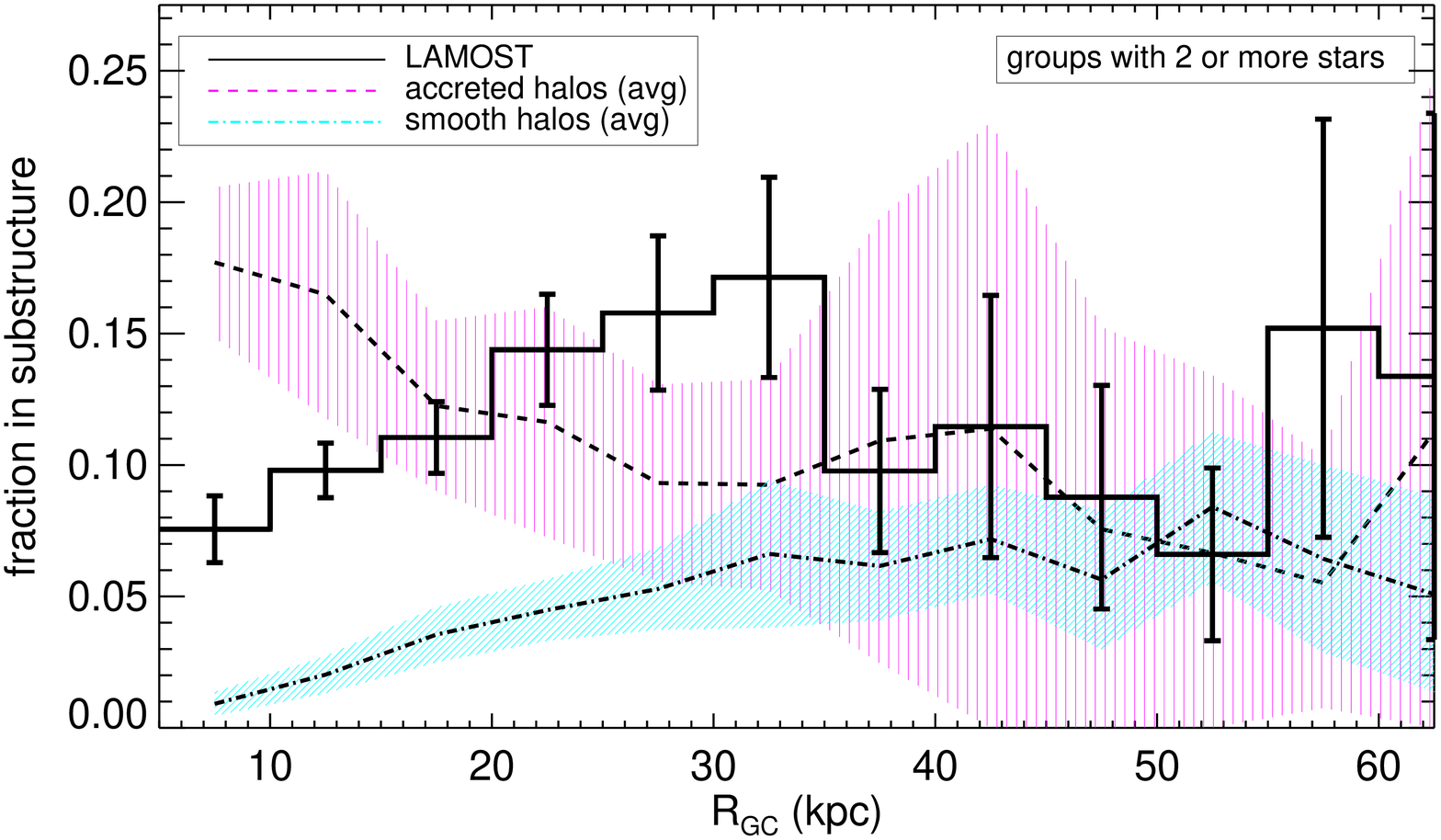}
\includegraphics[width=0.51\textwidth]{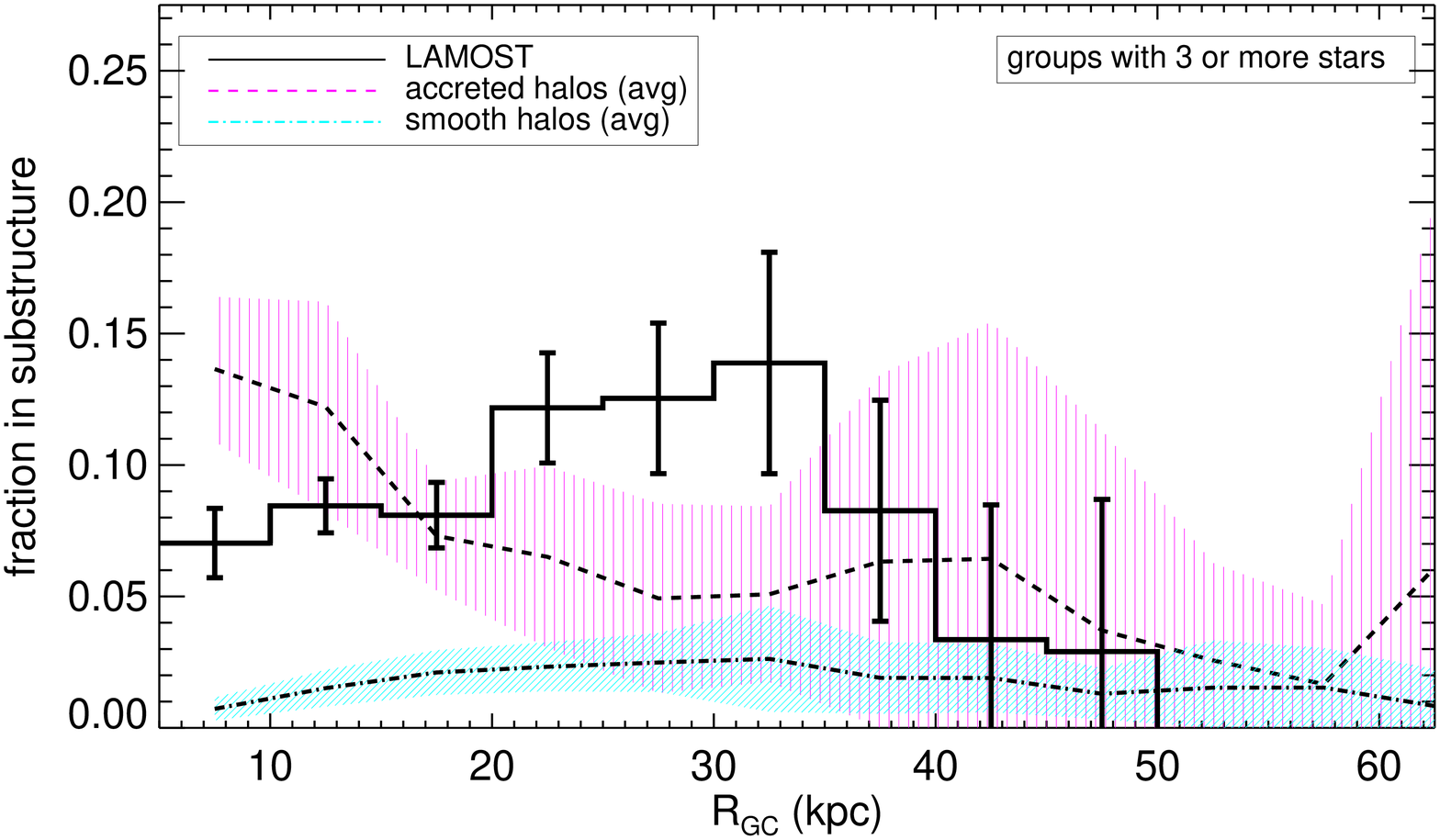}
\caption{Summary of results from Figure~\ref{fig:substruct_frac}. In the left panel, the solid histogram represents the fraction of stars in the LAMOST halo-giants sample that are part of SHARDS having 2 or more stars; the right panel shows SHARDS with 3 or more stars. Error bars represent Poisson uncertainties on the number counts in each bin. In both panels, the dashed line is the average from the 11 models of \citet{bj05}, with the magenta hashed region representing the 1$\sigma$ variation about this mean (this is the same as the gray filled regions in Fig.~\ref{fig:substruct_frac}). The dot-dashed line with cyan hashed region is the result from 10 simulated smooth halos (as in the right panels of Fig.~\ref{fig:substruct_frac}). This smooth halo fraction represents the level of false positives arising due to our SHARDS detection method.
The fraction of LAMOST stars in SHARDS with $\geq3$ stars is roughly $10\%$ at all radii $R_{\rm GC} \lesssim 40$~kpc. Beyond this radius, there are too few stars for the bins to be meaningful.}
\label{fig:substruct_frac_summary}
\end{figure*}

\section{Results and Discussion}\label{sec:results}

\subsection{Fraction of halo stars in substructure}\label{sec:substructfrac}

One of the main questions we wish to answer is what fraction of stars
in the Galactic halo reside in detectable phase-space
substructures. While the determination of an absolute fraction is
beyond the scope of this work, we can assess what fraction of the
LAMOST halo stars is part of substructure, and compare this to
expectations for what a similar survey would observe from the BJ05
mock halos.

Figure~\ref{fig:substruct_frac} shows the results of our substructure
search in the LAMOST DR1-3 data. Each of the panels in this figure
shows the fraction of observed stars that are identified as SHARDS in
each 5 kpc bin in Galactocentric radius. To compute this fraction, we
summed all stars in SHARDS, then subtracted the expected number of
stars in each bin. This fraction thus consists of the number of {\it
excess} stars in substructures. The upper row shows the fraction of 
all stars in bins containing two or more stars that are identified as excesses above three sigma,
and the lower panels require groups to have 3 or more stars for
inclusion. All panels show the observed fraction of substructure in
LAMOST as a solid black histogram. The panels on the left show results
from all 11 of the BJ05 halos as lines of different colors, with the
average and standard deviation as a dashed black line and shaded gray
region. On the right side, the dashed lines represent the average of
10 simulations of the smooth halo, with the gray shaded region
encoding 1$\sigma$ variation (where $\sigma$ is the standard
deviation) about this mean.

%

The upper-right panel of Figure~\ref{fig:substruct_frac} compares the
LAMOST observational results (solid black line) to the average of 10
smooth halo realizations (dashed line with 1$\sigma$ shaded
region), with bins required to have more than a single star in
each of the SHARDS. It is immediately clear that the stochastic sampling of the
halo by LAMOST yields some background of excess stars relative to the
Galaxia model predictions, but that the
likelihood of detecting multiple stars in a $>3\sigma$ excess in the
smooth halo is slim, yielding an average of less than 10\% of stars in
SHARDS over all radii. Thus, the fact that the LAMOST substructure
fraction (solid line) is larger than 10\% over nearly all radii less
than 50~kpc signals a significant difference between the smooth halo
and what we have observed. This fraction is consistent with the 10\%
lower limit on the fraction of Spaghetti survey data that were
estimated by \citet{shm+09} to be associated with halo substructures.



The upper-left panel of Figure~\ref{fig:substruct_frac} compares
SHARDS that contain multiple (i.e., $>1$) LAMOST halo stars to similar
detections in the 11 BJ05 model halos. From this figure we conclude
that, given the LAMOST selection function and our SHARDS algorithm,
the amount of substructure in the Galactic halo is consistent with
model halos built wholly of accreted satellites. Between $15 \lesssim
R_{\rm GC} \lesssim 60$~kpc, the fraction of LAMOST stars in SHARDS
follows closely the fraction from BJ05. Given that there is also an
excess above the smooth halo in this distance range, it is clear
that we are seeing a signal of substructure. The fraction of
substructure in the pure-accretion BJ05 models is found by our method
to be as low as $\sim10-20\%$ at all radii (on average); this arises
because of a combination of the LAMOST selection function and the fact
that debris from the earliest accretion events is mostly well-mixed at
present, and thus is no longer visible as four-dimensional
substructure by our method.

To isolate a sample with little expected contribution from false
positives, the bottom panels of Figure~\ref{fig:substruct_frac} show
the fraction of stars in substructure, but including only stars that
are in stellar excesses of 3 or more stars. As expected, the smooth
halo (lower-right panel) has very few detections of such groups, while
there are numerous SHARDS in LAMOST that contain $\geq3$ stars out to
$R_{\rm GC} \sim 40$~kpc (note that the lack of such stars beyond 40
kpc is likely simply due to the paucity of observed stars at such
distances). The $\gtrsim10\%$ fraction of LAMOST stars in
substructures with more than 3 stars exceeds the fraction derived from
running our algorithm on the BJ05 models (lower left panel) over radii
$20\lesssim R_{\rm GC} \lesssim 35$~kpc. This may be due to the
presence of the Sgr stream over much of the LAMOST footprint; few of
the BJ05 model halos contain such a massive, late-infalling satellite.

Figure~\ref{fig:substruct_frac_summary} summarizes our conclusions
from Figure~\ref{fig:substruct_frac}, focusing on the results for
SHARDS that contain 2 or more stars (left panel) and 3 or more stars
(right panel), and are thus securely identified as substructures with
little contamination from false positives. The solid black line in
both panels of Figure~\ref{fig:substruct_frac_summary} shows the
fraction of LAMOST stars that are part of SHARDS as a function of
Galactocentric radius. For comparison, the dot-dashed line shows
number of ``false positives'' as the average fraction of stars in
SHARDS from the 10 simulated smooth halos, with the 1$\sigma$
variation among models given by the cyan shaded region. The average of
the eleven BJ05 accreted halos is also shown as a dashed line, with
the variation among these mock halos shaded in magenta. The fraction
of stars in substructure in the LAMOST data exceeds that in the smooth
halo at all radii out to at least 40~kpc. In fact, the LAMOST results
exceed even the purely accreted halos at radii between $20 \lesssim
R_{\rm GC} \lesssim 35$~kpc, and the level of substructure is
consistent at nearly all radii. The fraction of LAMOST halo stars that
reside in SHARDS is consistently about 10\% above the floor set by the
false positives at all radii out to $R_{\rm GC} \lesssim
40$~kpc. Beyond 40~kpc, there are too few stars in our LAMOST catalogs
to draw strong conclusions.

\begin{figure}[!t]
\includegraphics[width=1.05\columnwidth]{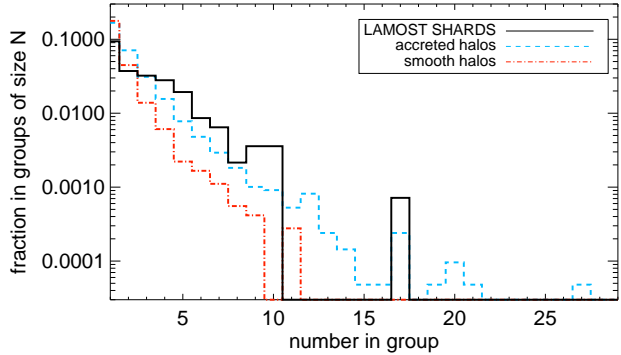}
\caption{Number of stars in SHARDS from LAMOST (solid line), BJ05 accreted-halo models (dashed blue line), and smooth-halo models (dot-dashed red line). The accreted and smooth-halo results include the concatenation of all of the models. To facilitate comparison, we show the bins as a fraction of all stars. As expected, LAMOST and the accreted halo have more large groups than the ``false positives'' from the smooth halo. Surprisingly, the SHARDS in LAMOST contain a larger fraction of groups with $3 < N < 10$ stars than the purely accreted halos.}
\label{fig:ngrp}
\end{figure}

In Figure~\ref{fig:ngrp} we show the background-subtracted group size
of SHARDS from LAMOST (black histogram), the accreted halos from BJ05
(blue dashed line), and the smooth Galaxia halos (dot-dashed red
histogram). As expected, there are more large groups ($N\gtrsim3$
stars) in the LAMOST and accreted-halo results than in the smooth halo
(i.e., the ``false positives''). Between $3 < N < 10$ stars per
group, the fraction of stars in SHARDS detected in LAMOST exceeds
even the fraction expected in the purely accreted halos. However,
there are more stars in groups with $>10$ stars in the BJ05 halos than
in LAMOST, suggesting that there may be some intrinsic difference
between the clustering scale of SHARDS and those predicted by models
of accretion-derived halos.

We conclude from analysis of Figures~\ref{fig:substruct_frac},
\ref{fig:substruct_frac_summary}, and \ref{fig:ngrp} that the
population of halo stars in LAMOST is clearly inconsistent with being
drawn from a smooth halo. Indeed, the fraction of Milky Way halo stars
in substructure is consistent with halos (from BJ05) created entirely
from accreted satellites. Our conservative choice requiring at least 3
stars in each SHARD places a lower limit of $\sim10\%$ of halo stars
in substructure at radii $R_{\rm GC} < 40$~kpc.

\subsection{Comparison of global properties of SHARDS to non-SHARDS halo stars}\label{sec:compare_properties}

\begin{figure}[!t]
\includegraphics[width=1.0\columnwidth]{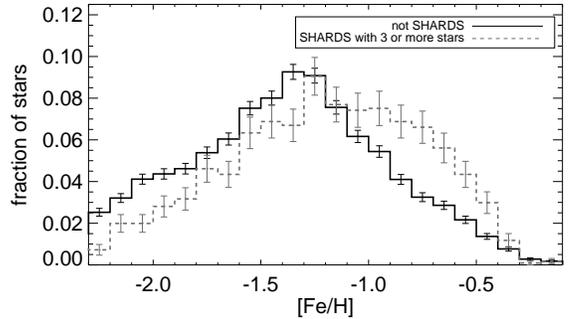}
\caption{Metallicities of stars {\it not} in SHARDS (solid black line) compared to those that are part of SHARDS with 3 or more stars (gray dashed line). Error bars are Poisson uncertainties on the number counts in each bin.
The stars that {\it are} part of substructure are on average more metal-rich than those that are not readily identified with substructure, as was also seen by \citet{srl+11} for ECHOS in SDSS. This suggests that the recent accretion history of the Milky Way has been dominated by luminous satellites.}
\label{fig:feh_shards_notshards}
\end{figure}

We now compare the properties of stars identified as part of SHARDS in
the LAMOST halo sample to those that are not members of
SHARDS. Figure~\ref{fig:feh_shards_notshards} shows the metallicity
distribution of all stars that are not part of SHARDS as a solid line;
members of SHARDS with 3 or more stars are shown as the dashed
histogram.\footnote{Note that this does not represent a true
metallicity distribution function (MDF) of the Galactic halo, because
we have not corrected for completeness or selection effects.} The
SHARDS are shifted toward higher metallicities than the stars that are
not in substructure, with a deficit of stars at [Fe/H]$\lesssim -1.6$
and an excess in all bins with [Fe/H]$> -1.1$ relative to the stars
that are not in SHARDS. \citet{srl+11} found a similar trend in the
ECHOS from SDSS; the mean metallicities of ECHOS are more metal-rich
than the average [Fe/H] of MSTO stars along the same line of
sight. The metallicity of debris is correlated with satellite
luminosity \citep{jbs+08}, such that more luminous satellites are more
enriched than their fainter counterparts. Furthermore, intact dwarf
galaxies in the Local Group exhibit a clear luminosity-metallicity
relation \citep{kcg+13}, with satellites at [Fe/H]$= -1.0$ at
luminosities of $L\sim10^8 L_\odot$. Because the substructures that
are observed as SHARDS (or ECHOS) result from relatively recent
accretion events, their bias toward more metal-rich populations
suggests that the late-infalling satellites contributing to the outer
Galactic halo have been predominantly luminous dwarfs rather than
metal-poor ultra-faint dwarf spheroidals. Of course, the most
prominent substructure in the halo -- the Sgr stream -- is an ongoing
accretion of a metal-enriched, luminous satellite, which may
contribute many of the metal-rich SHARDS we have
detected.\footnote{As a simple test of the number of Sgr stream stars contributing SHARDS to our study, we fit polynomials to the trends of distance and velocity with position from the compilation of Sgr observational data in Figure~6 of \citet{bke+14}. We then selected SHARDS within 15$^\circ$ of the Sagittarius plane, within $\pm30\%$ in distance from the polynomial trend, and less than 25 km~s$^{-1}$ in velocity from the polynomial fit to the Belokurov et al. data. This results in 43 SHARDS containing a total of 167 excess stars. This amounts to only $\sim8\%$ of the stars in SHARDS from our study. However, this is limited to the portions of the stream within which \citet{bke+14} presented data, and is thus likely missing some Sgr debris. We defer further detailed discussion of the Sgr stream as seen by LAMOST to later work.} Indeed, typical metallicities in the Sgr stream range from about $-1.2 < {\rm [Fe/H]} < -0.4$ (see summary in \citealt{lm16}), which is similar to the metallicity range of the excess stars in SHARDS.

\begin{figure}[!t]
\includegraphics[width=1.0\columnwidth]{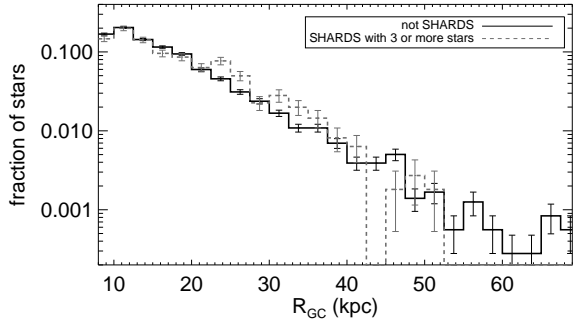}
\caption{Distribution of Galactocentric distances for stars {\it not} in SHARDS (solid black line) compared to those that {\it are} part of SHARDS having 3 or more stars (gray dashed lines). Error bars are Poisson uncertainties in the number counts in each bin.
Between $20 \lesssim R_{\rm GC} \lesssim 40$~kpc, the SHARDS density profile exceeds that of the underlying (not-SHARDS) sample. This may signal an increase in the importance of accretion vs. in situ halo stars at larger radii, as has been seen in simulations by, e.g., \citet{ans06}.}
\label{fig:rgc_shards_notshards}
\end{figure}

There is evidence for two components in the Galactic halo (the inner
and outer halo; \citealt{cbl+07, cbc+10, bci+12}), with a transition
at $R_{\rm GC} \sim 20$~kpc. In simulations of Milky Way-like
galaxies, \citet{ans06} showed that a break in the surface brightness
profile occurs naturally at $\sim20$~kpc from the galaxy center, where
within this radius most of the stars are formed in situ, and kicked up
during merger events, and outside 20~kpc, $\sim95\%$ of stars are
accreted. The accreted stars contribute excess luminosity beyond the
break relative to an extrapolation of the inner surface brightness
profile. In Figure~\ref{fig:rgc_shards_notshards}, we compare the
radial distribution of the ``smooth'' component in our LAMOST sample
(stars that are not in SHARDS) to stars that are in SHARDS. The SHARDS
(dashed line) deviate slightly from the ``not-SHARDS'' sample (solid
histogram) at radii of $\sim20-40$~kpc, with the SHARDS more prominent
in this range. While we have no way of distinguishing accreted vs. in
situ halo stars, this break in the relative numbers of SHARDS
vs. non-SHARDS stars suggests that accretion is a more important
contributor to the halo beyond $R_{\rm GC} \sim 20$~kpc than within
this radius.

\section{Conclusions}

We have presented a technique for the statistical identification of
substructure in the Galactic halo using spectroscopically-confirmed
RGB stars from LAMOST. Our method relies on comparison to the Galaxia
model \citep{sbj+11} along each line of sight, observing the model
halos with the LAMOST selection function to ensure a valid
comparison. We conservatively estimate that $\gtrsim10\%$ of the Milky
Way halo stars from LAMOST are part of substructures that we refer to
as SHARDS (Stellar Halo Accretion Related Debris Structures).  We
quantify the significance of our substructure detection by comparing
to a smooth-halo model (Galaxia) and to mock halos consisting entirely
of accreted satellites (from \citealt[BJ05]{bj05}). We find that the
fraction of halo stars in substructure exceeds what is expected from
statistical fluctuations due to incomplete sampling of a smooth halo
over all Galactocentric radii $R_{\rm GC} < 40$~kpc. Additionally, the
LAMOST substructure fraction follows closely the fraction of BJ05
stars that are identified with SHARDS, suggesting that the Galactic
halo is consistent with a purely accreted origin.

The SHARDS we have detected are in general more metal-rich than the
halo stars that are not part of substructure. The radial profile of
SHARDS differs from that of the non-SHARDS population beyond $R_{\rm
GC} \sim 20$~kpc, suggesting a break in stellar populations that may
be due to a transition from well-mixed early accretion debris and in
situ stars in the inner halo to late-infall accretion dominated
populations in the outer halo.

In this work, we are interested in the global characteristics of the
Galactic halo, but of course many known substructures are likely
present in our catalog of SHARDS. A brief examination shows evidence
of the Sgr stream, TriAnd, and the Virgo overdensity, and there are
likely many others within this data set. We defer discussion of the
previously known structures to later work, where we plan to examine
them in more detail. Likewise, a detailed comparison of the chemical
abundances of SHARDS to those of halo stars that are not in
substructure will come in future work. Currently, the LAMOST stellar
parameters pipeline derives only a bulk metallicity, [Fe/H], but not
detailed abundances such as [$\alpha$/Fe] or [C/Fe]. Lee et al. (2015)
have shown that quality measurements of these abundances can be
readily achieved from LAMOST spectra, so we expect in the near future
to analyze the abundance signatures of SHARDS, as was done by, e.g.,
\citet{srl+12} for ECHOS from SDSS. Chemical abundances
may also provide an additional dimension that can be used to distinguish
accretion relics from in situ halo populations. As the LAMOST survey
continues to fill more contiguous area to more uniform depths, it will
be possible to characterize the level of substructure on different
spatial scales, with which we can explore the luminosity function and
orbit types of infalling satellites that have contributed to the halo
\citep{jbs+08}.

Our method to identify substructure is similar in many respects to
other techniques that have been used (e.g., the {\it 4distance} --
\citealt{shm+09, jmm+16}; ECHOS -- \citealt{sra+09}; two-point
correlation function: \citealt{ccf+11, xry+11}). Our technique takes
advantage of the large filling factor of the LAMOST survey on the sky,
and we account for the complicated selection function of the
survey. We conclude that, beyond $R_{\rm GC} > 20$~kpc, the fraction
of halo giants from LAMOST that are in substructure is consistent with
expectations from stellar halos \citep{bj05} built entirely from
accreted satellites.

\acknowledgements

We thank the referee for valuable comments that greatly improved this work. J.L.C. is grateful for the hospitality and support from Chao Liu and
Licai Deng during his summer in Beijing, and for the Chinese Academy
of Sciences President's International Fellowship that supported this
work.  This work was supported by the U.S. National Science Foundation
under grants AST 09-37523 and AST 14-09421, and by The Marvin Clan,
Babette Josephs, Manit Limlamai, and the 2015 Crowd Funding Campaign
to Support Milky Way Research. C.~L. also acknowledges the Strategic
Priority Research Program ``The Emergence of Cosmological Structures''
of the Chinese Academy of Sciences, grant No. XDB09000000, the
National Key Basic Research Program of China, grants No. 2014CB845700,
and the National Science Foundation of China, grants No. 11373032 and
11333003. T.C.B. acknowledges partial support for this work from
grants PHY08-22648; Physics Frontier Center/Joint Institute or Nuclear
Astrophysics (JINA), and PHY 14-30152; Physics Frontier Center/JINA
Center for the Evolution of the Elements (JINA-CEE), awarded by the US
National Science Foundation.  Guoshoujing Telescope (the Large Sky
Area Multi-Object Fiber Spectroscopic Telescope; LAMOST) is a National
Major Scientific Project built by the Chinese Academy of
Sciences. Funding for the project has been provided by the National
Development and Reform Commission. LAMOST is operated and managed by
the National Astronomical Observatories, Chinese Academy of
Sciences. This publication makes use of data products from the Two
Micron All Sky Survey, which is a joint project of the University of
Massachusetts and the Infrared Processing and Analysis
Center/California Institute of Technology, funded by the National
Aeronautics and Space Administration and the National Science
Foundation.

This research made use of the publicly available software Galaxia and
Topcat \citep{t05}, the IDL Astronomy Library \citep{l93}, and NASA's
Astrophysics Data System. This work relied on the {\it zen} computing
cluster at the National Astronomical Observatories of China, Chinese
Academy of Sciences in Beijing. We thank Ricardo Mu{\~n}oz for
valuable discussion of an early draft of this work.

\bibliographystyle{apj}

\end{document}